\documentclass[aps,prb,amsmath,superscriptaddress,amssymb,longbibliography,twocolumn]{revtex4-2}
\usepackage{graphicx}
\usepackage{physics}
\usepackage{amsmath}
\usepackage{wasysym}
\usepackage{graphicx}
\usepackage{gensymb}
\usepackage{braket}
\usepackage{natbib}
\usepackage{comment}
\usepackage{physics}
\usepackage{color}
\usepackage[normalem]{ulem}
\usepackage{setspace}
\usepackage{booktabs}
\usepackage{tabularx}
\usepackage{quantikz}
\usepackage[colorlinks=true,citecolor=blue,linkcolor=magenta, breaklinks=true]{hyperref}

\begin{document}
	\title{Basis Adaptive Algorithm for Quantum Many-Body Systems on Quantum Computers}

\begin{abstract}

We introduce a Basis Adaptive (BA) algorithm for hybrid quantum-classical
simulation of correlated quantum many-body systems.
Starting from a small set of physically motivated bitstrings, the
algorithm iteratively applies a single-step first-order Trotterized
circuit on a quantum processor, filters the sampled configurations by
enforcing $U(1)$ spin conservation and lattice reflection symmetry, and
classically diagonalizes the Hamiltonian in the resulting reduced
Hilbert space.
This design avoids the variational optimization overhead of VQE, the
deep coherent circuits required by QPE, and the symmetry-violating
subspaces that arise in SKQD.
The ground-state energy error is bounded analytically by
$\sqrt{8}\,\|H\|(1-\sqrt{\alpha_{D_T}})^{1/2}$, where $\alpha_{D_T}$ is the
probability weight captured by the $D_T$ sampled basis states; this bound
connects algorithm performance directly to ground-state sparsity and
explains the observed accuracy hierarchy across phases.
Benchmarked on the spin-$\frac{1}{2}$ Heisenberg XXZ chain ($N$ up to $62$ qubits,
IBM Heron processor), the algorithm achieves $3.5\%$ energy error in the
gapped N\'{e}el phase ($\Delta=2.0$)
%compared to $7.28\%$ for SKQD at
%the same subspace dimension — 
and below $0.5\%$ at the ferromagnetic
boundary ($\Delta=-1.0$).
Accuracy degrades to $28.7\%$ in the strongly quasi-long-range-ordered
regime ($\Delta=0.5$), 
%consistent with the logarithmically slow
%convergence of $\alpha_L$ predicted by Luttinger-liquid theory.
Spin–spin correlation functions are reproduced %high fidelity 
across all regimes, confirming that symmetry-filtered real-time sampling offers
a practical and noise-resilient pathway to ground-state properties on
near-term quantum hardware.

%\textcolor{red}{
%A new basis adaptive algorithm for hybrid quantum-classical platforms is introduced to efficiently find the ground-state (gs) properties of quantum many-body systems. The method addresses limitations of many algorithms, such as Variational Quantum Eigensolver (VQE) and Quantum Phase Estimation (QPE) etc by using shallow Trotterized circuits for short real-time evolution on a quantum processor. The sampled basis is then symmetry-filtered by using various symmetries of the Hamiltonian which is then classically diagonalized in the reduced Hilbert space. We benchmark this approach on the spin-1/2 XXZ chain up to 24 qubits using the IBM Heron processor. The algorithm achieves sub-percent accuracy in ground-state energies across various anisotropy regimes. Crucially, it outperforms the Sampling Krylov Quantum Diagonalization (SKQD) method, demonstrating a substantially lower energy error for comparable reduced-space dimensions. This work validates symmetry-filtered, real-time sampling as a robust and efficient path for studying correlated quantum systems on current near-term hardware.}
\end{abstract}

\author{Anutosh Biswas}
\affiliation{S.N. Bose National Centre for Basic Sciences, Kolkata 700106, India.}
\author{Sayan Ghosh}
\affiliation{S.N. Bose National Centre for Basic Sciences, Kolkata 700106, India.}
\author{Ritajit Majumdar}
\affiliation{IBM Quantum, IBM India Research Lab, Bengaluru, India.}
\author{Mostafizur Rahaman Laskar}
\affiliation{IBM Quantum, IBM India Research Lab, Bengaluru, India.}
\author{Nicholas Bronn}
\affiliation{IBM Quantum IBM T.J.
Watson Research Center, Yorktown Heights, NY, USA.}
\author{Manoranjan Kumar}
\email{manoranjan.kumar@bose.res.in}
\affiliation{S.N. Bose National Centre for Basic Sciences, Kolkata 700106, India.}
\date{\today}

\maketitle	
%%%%  INTRODUCTION    %%%%%%

\section{Introduction}

Quantum many-body effects in condensed matter systems are ubiquitous and play a crucial role in giving rise to a variety of emergent quantum phases, including unconventional magnetism \cite{balents2010spinliquids, moessner2006frustration,gardner2010pyrochlore}, superconductivity \cite{lee2006htsc,scalapino2012unconventional,keimer2015review}, quantum spin liquids \cite{savary2017qsl, zhou2017qsl,yan2011kagome}, and entanglement-driven orders  \cite{anderson1987resonating, ImadaRMP}. However, studying the quantum many body systems remains intractable due to the exponential growth of the Hilbert-space dimension with system size. This exponential scaling makes exact solutions infeasible for most systems, while exactly solvable small systems suffer from significant finite-size effects, which hinder a true understanding of real materials \cite{mah00, Fetter, sandvik2010computational}. 
Standard numerical approaches—such as exact diagonalization (ED)\cite{lin1990exact,nataf2014exact,sandvik2010computational}, quantum Monte Carlo \cite{foulkes2001quantum,pollet2012recent,gubernatis2016quantum,sandvik1991quantum,ceperley1986quantum,santos2003introduction}, and tensor-network methods \cite{montangero2018introduction,banuls2023tensor,ran2020tensor,fishman2022itensor}—have provided important insights, but each suffers from inherent limitations. ED is restricted by small system sizes, quantum Monte Carlo is limited by the fermionic and frustrated sign problems \cite{pan2022sign,troyer2005computational}, and tensor-network methods struggle in regimes with strong entanglement or long correlation lengths \cite{SchollwockDMRG}.

%\textcolor{red}{Quantum many-body systems remain difficult to analyze because the Hilbert-space dimension grows exponentially with system size. This rapid growth makes exact solutions infeasible for all but the smallest systems and complicates the study of strongly correlated phases such as unconventional magnetism, critical quantum liquids, and entanglement-driven orders \cite{anderson1987resonating, ImadaRMP}. Standard numerical approaches—including exact diagonalization\cite{lin1990exact,nataf2014exact,sandvik2010computational}, quantum Monte Carlo\cite{foulkes2001quantum,pollet2012recent,gubernatis2016quantum,sandvik1991quantum,ceperley1986quantum,santos2003introduction}, and tensor-network methods\cite{montangero2018introduction,banuls2023tensor,ran2020tensor,fishman2022itensor} have provided major insights, but each faces inherent limitations. Exact diagonalization is constrained by system size, quantum Monte Carlo encounters the fermionic and frustrated-sign problems \cite{pan2022sign,troyer2005computational}, and tensor-network methods struggle in regimes with large entanglement or long correlation lengths \cite{SchollwockDMRG}.}

Quantum computation offers a potential path forward by encoding many-body wavefunctions directly in quantum hardware. A variety of algorithms have been proposed for ground-state preparation and Hamiltonian simulation. The Variational Quantum Eigensolver (VQE) \cite{tilly2022variational,kandala2017hardware,liu2019variational,zhang2022variational,peruzzo2014variational} is attractive for near-term hardware devices because it requires shallow circuits, but its performance is limited by optimization landscapes that become highly nonconvex for strongly correlated states \cite{mcclean2018barren}. Quantum Phase Estimation (QPE) \cite{KitaevQPE, nielsen2000qci} can achieve high precision but demands deep, coherent circuits that are only practical on fault-tolerant quantum processors. Krylov-subspace methods \cite{yoshioka2025krylov}, including Sample-based Krylov Quantum Diagonalization (SKQD) \cite{yu2025quantum,brooks2026ground}, mitigate some depth requirements but often generate noisy or symmetry-violating subspaces, reducing accuracy. Imaginary-time evolution schemes \cite{MottaITE, McArdleITE} involve effective nonunitary dynamics that can be approximated with ancilla-based circuits and are sensitive to noise. Real-time evolution approaches \cite{ChildsTrotter, CampbellTrotter} can be implemented using shallow Trotter circuits, but short-time evolution may not capture the dominant ground-state structure unless basis states are carefully selected and filtered. 
A complementary class of hybrid quantum-classical approaches seeks to identify a compact low-energy subspace using quantum sampling, followed by classical diagonalization within the sampled subspace. Early work in this direction introduced Quantum-Selected Configuration Interaction (QSCI)\cite{kanno2023quantum,mikkelsen2025quantum,sugisaki2025hamiltonian}, in which measurement outcomes from a quantum computer are used to select the most relevant basis configurations for a configuration-interaction calculation \cite{kanno2023quantum}. Subsequently, adaptive variants such as ADAPT-QSCI \cite{nakagawa2024adapt} were developed to iteratively improve the sampling state. More recently, time-evolved states have been employed to enhance the quality of the sampled subspace in TE-QSCI \cite{mikkelsen2025quantum}, while related developments include HSB-QSCI \cite{sugisaki2025hamiltonian} and Sample-based Krylov Quantum Diagonalization  (SKQD) \cite{yu2025quantum}, which construct effective subspaces through quantum-generated samples. These studies demonstrate the potential of quantum-assisted subspace construction as a promising route toward solving quantum many-body problems beyond the reach of ED. Despite these advances, constructing a compact basis that accurately captures the low-energy physics remains challenging, particularly as the Hilbert-space dimension grows with the system size and the sampling efficiency becomes increasingly important ~\cite{patra2025physics}.

These challenges motivate the need for a systematic and hardware-efficient framework for constructing compact, physically meaningful basis sets. Here, we introduce a Basis Adaptive (BA) algorithm in which a reduced Hilbert space is generated through short duration real-time evolution of physically motivated computational-basis states on a quantum processor, thereby capturing the essential low-energy physics while avoiding the exponential cost of the full Hilbert space. In contrast to existing QSCI- and Krylov-based approaches, the generated configurations are explicitly filtered to preserve the symmetries of the Hamiltonian, including conservation of total $S^z$ and lattice reflection symmetry, before the reduced Hamiltonian is diagonalized classically. The iterative selection of the most probable basis states further concentrates the subspace onto the physically relevant low-energy sector. Because the method naturally focuses on symmetry-allowed low-energy configurations, it avoids the variational optimization overhead of VQE, the deep circuits required by QPE, and reduces the sampling of symmetry-violating states that can arise in generic Krylov-based constructions.
%In this work, we introduce an algorithm in which reduced Hilbert space with only relevant bases are obtained using an efficient time evolution on a quantum computer. 
%in which a reduced Hilbert space is built from measurement outcomes obtained after a short real-time evolution of an initial trial state.
%These bases are explicitly filtered to preserve key symmetries of the Hamiltonian, including conservation of total  $S^{z}$ and lattice reflection symmetry. The Hamiltonian is then diagonalized within this reduced Hilbert space to obtain the ground-state energy and wavefunction. Because the method naturally focuses on the low-energy sector, it avoids the heavy optimization overhead of VQE, the deep circuits required by QPE, and the noise-induced instabilities that may arise in SKQD. Its reliance on shallow Trotterized circuits also makes it practical for near-term hardware.

In this work we study the spin-$\frac{1}{2}$  Heisenberg XXZ chain to evaluate the performance of the algorithm. The XXZ model is an interesting model which exhibits a ferromagnetic phase $\Delta < -1$ , a critical Tomonaga–Luttinger liquid for $-1 < \Delta \leq 1 $, and an antiferromagnetic Néel-ordered phase for  $\Delta > 1 $ \cite{GiamarchiBook, TakahashiBook}. These regimes differ markedly in their entanglement structure and quantum fluctuation strength, making the model an ideal benchmark. We implement the BA algorithm on IBM's Heron quantum processor for systems containing up to 62 qubits and compare the results with those obtained from Density Matrix Renormalization Group (DMRG) and SKQD.

Our results demonstrate that the proposed approach achieves sub-percent accuracy in ground-state energies across a broad range of anisotropy values. It also reproduces spin–spin correlation functions with high fidelity and outperforms SKQD for comparable reduced-space dimensions. These findings indicate that symmetry-filtered, real-time sampling provides a robust and efficient pathway for studying correlated quantum systems on currently accessible quantum hardware.

This paper is organized into six sections. In Section II, we introduce the model Hamiltonian and the numerical approaches used to benchmark our results. Section III presents the BA algorithm employed for the quantum computations. In Section IV, we discuss the energy error bounds and ground-state sparsity. Section V contains the numerical and quantum-computing results. Finally, in Section VI, we summarize our findings and provide concluding remarks.
   
\section{Model Hamiltonian and numerical approach}
The spin-\(\tfrac{1}{2}\) XXZ chain is described by the Hamiltonian 
\begin{equation}\label{eq1}
H = J \sum_{j=1}^{L} \left( S_j^{x} S_{j+1}^{x} + S_j^{y} S_{j+1}^{y} + \Delta\, S_j^{z} S_{j+1}^{z} \right),    
\end{equation}
where $S_j^{\alpha}=\frac{1}{2}\sigma_j^{\alpha}$ are spin-$1/2$ operators, and $J=1$ sets the energy scale and antiferromagnetic nature of exchange, $\Delta$ as the anisotropy parameter controlling the interaction along the z-axis. 
In this work, we employ DMRG to benchmark the results obtained from 
the BA algorithm  for system sizes up to $N=62$ \cite{sandvik}. Using these complementary numerical approaches, we compare the ground state (gs) energy, spin–spin correlation functions obtained from the BA algorithm and ED method for system size up to $N=24$.  These comparisons serve as benchmarks for assessing the accuracy and performance of the BA Algorithm.

\section{Basis adaptive Algorithm}

For a given quantum system, the ground-state wavefunction can always be expressed 
as a linear combination of the computational basis states of the Hilbert space. 
If $\{ \ket{b_0},  \ket{b_1} ..\ket {b_k} ...\}$ denotes a complete basis, then the ground state may be written as
\begin{equation}
    |\phi_0\rangle = \sum_{k} c_k\, |b_k \rangle .
\end{equation}

In most physical systems, only a very small fraction of these basis states 
contribute significantly to the ground-state wavefunction, while the majority 
have negligible amplitudes i.e, $c_k \approx 0$. Therefore, we can construct a trial wavefunction using the most probable basis states from the ground state,
%Therefore, it is natural to construct an approximate ground state using only the most probable basis states,
\begin{equation}
    |\phi_{trial}\rangle \approx 
    \sum_{k \in \mathcal{S}}^{N_{SD}} c_{k}'\, |b_k\rangle ,
\end{equation}
where $\mathcal{S}$ denotes the subset of basis states with significant finite amplitudes $c_{k}'$, and $N_{SD}$ is the \emph{subspace dimension}, i.e., the number of basis states retained in the truncated subspace.
Our final goal is to get the approximated ground state wavefunction $|\tilde\phi\rangle$, therefore,  we begin with a very small set  dominant basis states $\{|b_k\rangle\}$ and evolve them through a short time step,
$\Delta t$. The short-time evolution of the approximate state can be written as 
\begin{equation}
    |\phi_{trial}^{\prime}(\Delta t)\rangle 
    = e^{-i H \Delta t}\, \sum_{k}c_{k}' \, |b_k\rangle .
\end{equation}
The time evolution is unitary operator and for small $\Delta t$, Taylor expansion of the time evolution operator has the following form,
\begin{equation}
    e^{-i H \Delta t} 
    \approx I - i\Delta t\, H 
    + \frac{(i\Delta t\, H)^2}{2}
    + \mathcal{O}(\Delta t^3),
\end{equation}
After a short-time $\Delta t$ approximate state can be written as
\begin{equation}
    |\phi_{trial}^{\prime}(\Delta t)\rangle 
    \approx 
    \left[
        I - i\Delta t\, H 
        + \frac{(i\Delta t\, H)^2}{2}
        + \cdots
    \right]
    \sum_{k} c_{k}' \, |b_k\rangle .
\end{equation}

The evolved state may also be expanded in the exact eigenbasis 
$\{|\psi_n\rangle\}$ of the Hamiltonian,
\begin{equation}
    |\phi_{trial}^{\prime}(\Delta t)\rangle 
    = \sum_{n} a_{n} \, e^{-iE_n \Delta t}\, |\psi_n\rangle ,
\end{equation}
and re-expanding it back in the computational basis yields
\begin{equation}
    |\phi_{trial}^{\prime}(\Delta t)\rangle 
    = \sum_{b_k' \in S_1}^{N^{1}_{SD}} a_{k}'\, |b_k'\rangle .
\end{equation}

The long time evolution naturally generates many additional basis states and the evolved wavefunction becomes function of various basis states which are generated through two-, four- or higher order spin flips, expanding the size of the truncated Hilbert space. To get a flavor of this we used spin flipped basis states for obtaining accurate ground state of models in Ising limit of the Hamiltonian starting with antiferromagnetic Ising basis and the calculation is presented in the Appendix~\ref{app:trotter}. It is shown that how the transverse anisotropy can be treated perturbatively and second and fourth order of contribution are just two spin flip or four spin flip terms. However, higher order of spin-flip process may generate all the basis states and these states may not be specifically relevant to only low-energy eigen-states of the Hamiltonian.  

In this work, we are interested only in the ground state, thus, we want to generate only those basis states of the Hilbert space which are relevant to the ground state. We follow an iterative algorithm \cite{rahaman2023machine}, where, we start with only a few basis states which are most relevant for the ground state of the Hamiltonian and the operation of the Hamiltonian on these basis states are very similar to the time evolution which generates many newly configuration. We now construct the Hamiltonian matrix in the newly expanded reduced Hilbert space and get ground state wavefunction. The most probable basis states, i.e states with highest probability, are selected and these basis states are time evolved to form the expanded Hilbert space. This process is repeated till the gs energy is converged. The overall workflow of the basis-adaptive algorithm is summarized in Fig.~\ref{fig:QC_Algorithm}.

This method can be efficiently implemented on the Quantum Computer (QC) and basis states can be evolved with short time using the QC and the construction of the Hamiltonian matrix and gs wavefunction calculation is done on classical computer. In this work we develop a basis adaptive algorithm which used hybrid quantum and classical  approach to solve the quantum many body model systems. The steps of the BA algorithm can be described below:  
\begin{enumerate}
    \item \textbf{Initialization:}  
We prepare a small set of representative bitstrings, 
$\{b_k^{(0)}\}$, which encode the dominant spin configurations contributing to the ground state of the Hamiltonian $H$. 
Each bitstring corresponds to a computational-basis state $\ket{b_k^{(0)}}$ within the Hilbert space of the system $H$. 
Unlike conventional wavefunction-based approaches, we treat each $\ket{b_k^{(0)}}$ as an independent sample wavefunction. 
The initial configurations can be chosen from the ground state of the Ising (axial) part of $H$ or guided by the most probable basis states obtained from solution of small system size. 
For example, in the XXZ Hamiltonian (Eq.~\ref{eq1}) with $J>0$ and $\Delta \ge 1$, the two relevant configurations are
\[
\ket{b_1^{(0)}} = \ket{\uparrow\downarrow\uparrow\downarrow\cdots}, \quad
\ket{b_2^{(0)}} = \ket{\downarrow\uparrow\downarrow\uparrow\cdots}.
\]

\item \textbf{Time Evolution:}  
In the $i$-th iteration, each configuration $\ket{b_k^{(i)}}$ is evolved independently under the unitary operator for short time evolution 
$U(\Delta t) = e^{-iH\Delta t}$:
\begin{equation}
e^{-iH\Delta t}\ket{b_k^{(i)}}
=
\sum_{\ket{b_j}\in\mathcal{B}_{k}^{(i)\prime}}
c_{jk}^{(i)}\,\ket{b_j},
\end{equation}
where $\mathcal{B}_{k}^{(i)\prime}$ denotes the set of computational basis states generated.

For the XXZ Hamiltonian,
\begin{align}
    H &= \sum_j H_j, \nonumber \\
    H_j &= \tfrac{1}{2}\left(S_j^+S_{j+1}^- + S_j^-S_{j+1}^+\right) + \Delta S_j^z S_{j+1}^z.
\end{align}
Using the first-order Lie--Trotter decomposition,
\begin{equation}
    e^{-iH\Delta t} \approx \prod_j e^{-iH_j\Delta t} + \mathcal{O}((\Delta t)^2),
\end{equation}
each local term $H_j$ acts on two neighboring spins, allowing the evolution to be implemented through a sequence of two-qubit unitaries on a quantum circuit (see Fig.~\ref{fig:trotter_xxz}). The detailed construction of the circuit is discussed in the Appendix~\ref{app:trotter}. The same circuit is applied independently to each input bitstring $\ket{b_k^{(i)}}$, producing the evolved configurations $\ket{b_k^{(i)'}}$.

\begin{figure}[h]
\includegraphics[width=1.0\columnwidth]{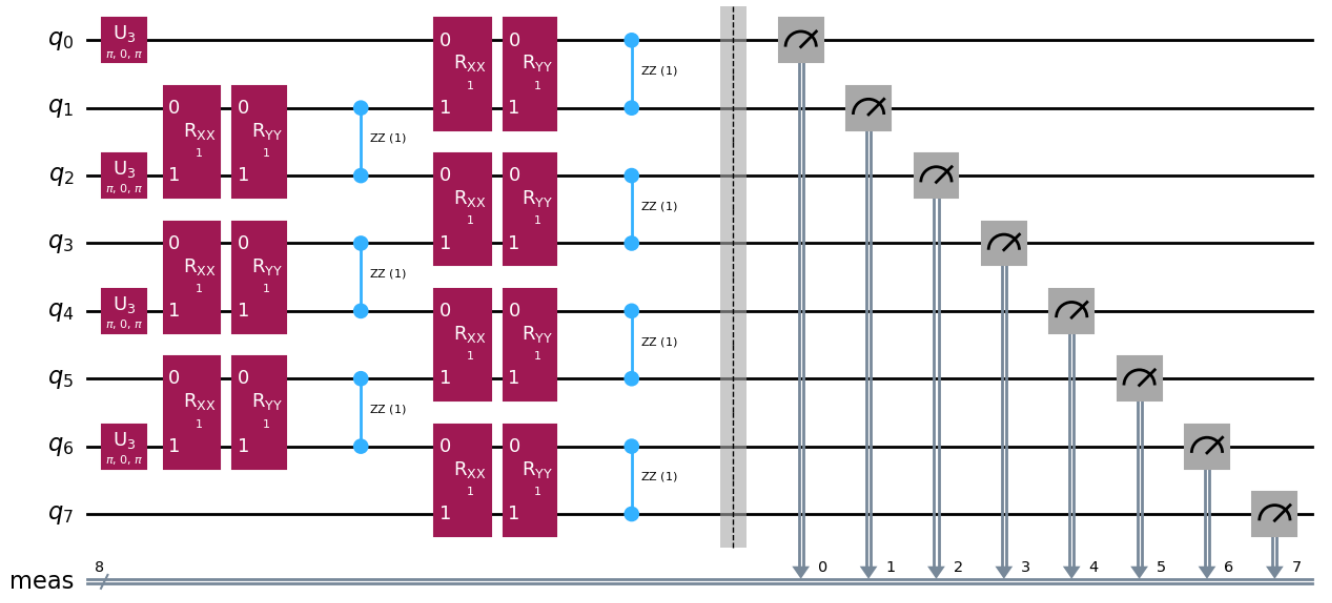}
\caption{Trotterized quantum circuit for the XXZ Hamiltonian in Eq. \ref{eq1} based on the first-order \textbf{Lie--Trotter decomposition}. 
    The bond exchange of Hamiltonian $H_k = J (X_k X_{k+1} + Y_k Y_{k+1}) + \Delta Z_k Z_{k+1}$ is decomposed into non-commuting parts 
    $H_{XY}$ and $H_{ZZ}$, whose time evolutions are implemented sequentially using $R_{XX}$, $R_{YY}$, and $ZZ$ rotation gates. 
    The single-qubit $U_3$ gates prepare the initial state, and the layered structure of $R_{XX}$, $R_{YY}$, and $ZZ$ gates realizes 
    one Trotter step of the evolution.}
    \label{fig:trotter_xxz}
\end{figure}

\item \textbf{Measurement, Union, and Symmetry Filtering:}  
In the third step, each evolved configuration $\ket{b_k^{(i)'}}$ is measured in the computational basis to obtain a distribution $\mathcal{B}_{k}^{(i)'}$. 
The complete set of bitstrings generated from all evolved configurations is then combined to form their union, 
\[
\mathcal{B}^{(i)'} = \bigcup_k \{b_k^{(i)'}\}.
\]
From this unified set $\mathcal{B}^{(i)'}$, we retain only those bitstrings that satisfy the required physical constraints — specifically, conservation of the total $S^z$ component and the $U(1)$ symmetry of the Hamiltonian. 
Furthermore, we symmetrize the set by including partners of bitstrings representing configurations related by the reflection invariably of the Hamiltonian.
The resulting symmetry-adapted set of bitstrings is denoted as $\mathcal{B}^{sym(i)'}$, ensuring that the reduced subspace preserves both $U(1)$ and reflection symmetries.

\item \textbf{Hilbert-Space Reconstruction and Classical Diagonalization:}  
In the fourth step, the filtered and symmetrized bitstrings $\mathcal{B}^{sym(i)'}$ define the reduced Hilbert space for the next iteration. 
Within this reduced subspace, we construct the Hamiltonian matrix corresponding to $H$ and perform an ED calculation on a classical computer. The lowest eigenvalue and the associated eigenvector, denoted as $E^i$ and $\ket{\Psi^i}$, respectively, represent the approximate ground-state energy and wavefunction at the $i$-th iteration.

    \item  In the fifth step, if our ground state energy is converged then we calculate all the relevant quantity like correlation functions, spin density etc and exit or we construct updated wavefunction retaining $m_i$ number of basis states which have highest probability $c^2_i$ in the ground state wavefunction of the projected Hamiltonian from ED calculation. These most probable bases \{$b^{i+1}_{m}$\} work as initial basis set for the next iteration $(i+1)^{th}$ and repeat the step two-five till the energy is converged. Typically, the dominant basis states contributing to the ground state are recovered within two iterations, leading to rapid convergence of the ground-state energy.
\end{enumerate}

\begin{figure}[h]
\includegraphics[width=1.0\columnwidth]{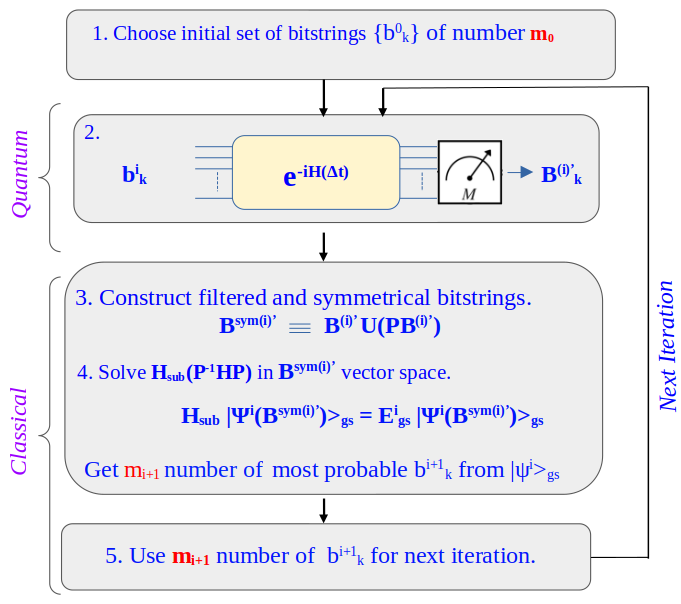}
%\caption{Schematic representation of one complete iteration of the proposed quantum--classical hybrid algorithm. The process begins with an initial state $\psi_0$, which is time-evolved under the Hamiltonian $\mathcal{H}$ to obtain  $\psi_t = e^{iH\Delta t}\psi_0$. The time-evolved states are measured to generate bitstrings (sampling), from which a reduced Hamiltonian is constructed and diagonalized to obtain the ground-state energy $E_0[\psi_t]$. The most probable basis states are then selected as new initial states for the next iteration. Throughout the process, the algorithm ensures that all sampled and reconstructed basis states preserve the $U(1)$ spin-rotation symmetry about the $z$-axis and the reflection symmetry.}
\caption{Schematic workflow of the Basis Adaptive Algorithm. An initial set of $m_{0}$ bitstrings is evolved using the Trotterized time-evolution operator on a quantum device and measured to generate a new set of sampled bitstrings. These bitstrings are then expanded by adding symmetry-related configurations, forming a symmetrized subspace in which the Hamiltonian is classically diagonalized. The $m_{i+1}$ most probable bitstrings from the resulting eigenstate are selected and used to define the basis set for the next iteration.}
\label{fig:QC_Algorithm}
\end{figure}

%============================================================
\section{Energy Error Bounds and Ground-State Sparsity}
\label{sec:bounds}
%============================================================

A central quantity governing the accuracy of any subspace diagonalization
method is how well the chosen reduced Hilbert space captures the true
ground state $|\phi_0\rangle$.  We characterize this through the
sparsity parameter $\alpha_{D_T}$, the total probability weight
carried by the $D_T$ largest computational-basis coefficients of
$|\phi_0\rangle$~\cite{yu2025quantum}.

%----------------------

\subsection{General energy bound}
\label{sec:general_bound}

To understand how the number of retained bitstrings controls the
accuracy of the BA method, we derive an upper bound on the energy
error following the analysis of Refs.~\cite{yu2025quantum,RobMottaSQD}.

Let the exact ground state be expanded in the computational basis as

\begin{equation}
|\phi_0\rangle
=
\sum_{j=1}^{D_{\mathcal H}}
c_j |b_j\rangle ,
\end{equation}

where the coefficients are sorted in descending order,

\begin{equation}
|c_1| \ge |c_2| \ge \cdots \ge |c_{D_{\mathcal H}}|.
\end{equation}

Here $D_{\mathcal H}$ denotes the full Hilbert-space dimension.
Retaining a number of largest-amplitude basis states $D_T$, we define the
cumulative retained weight

\begin{equation}
\alpha_{D_T}
=
\sum_{j=1}^{D_T}|c_j|^2 .
\label{eq:alphaL_def}
\end{equation}

The BA approximation to the ground state is then

\begin{equation}
|\tilde\phi\rangle
=
C^{-1}
\sum_{j=1}^{D_T}
c_j |b_j\rangle ,
\end{equation}

with normalization

\begin{equation}
C
=
\left(
\sum_{j=1}^{D_T}|c_j|^2
\right)^{1/2}
=
\sqrt{\alpha_{D_T}}.
\end{equation}

Defining

\begin{equation}
|\phi'\rangle
=
|\tilde\phi\rangle-|\phi_0\rangle ,
\end{equation}

and applying the Cauchy--Schwarz inequality (see
Appendix~\ref{app:bounds}), the energy error is bounded by

\begin{equation}
\boxed{
\Delta E
\le
\sqrt{8}\,\|H\|
\left(
1-\sqrt{\alpha_{D_T}}
\right)^{1/2}
}
\label{eq:master_bound}
\end{equation}

where $\|H\|$ is the spectral norm of the Hamiltonian.

The bound depends entirely on the retained probability weight
$\alpha_{D_T}$. Consequently, the convergence of the BA method is
determined by how rapidly the sorted coefficient distribution
$|c_j|^2$ decays with the rank index $j$. We now analyze this behavior
for the different phases of the XXZ chain.

\subsection{Short-range / gapped regime ($\Delta > 1$)}
\label{sec:gapped}

In the N\'eel phase ($\Delta>1$), the ground-state wavefunction is
strongly localized in the computational basis. Numerically, we find
that the sorted probabilities are well described by an exponential
form,

\begin{equation}
|c_j|^2
\simeq
c_0 e^{-j/j_0},
\label{eq:gapped_coeff}
\end{equation}

where $j_0$ is a characteristic decay scale.

As shown in Appendix~\ref{app:qlro}, this exponential form implies

\begin{equation}
1-\sqrt{\alpha_{D_T}}
\simeq
\frac12 e^{-D_T/j_0}.
\end{equation}

Substituting into Eq.~\eqref{eq:master_bound} yields

\begin{equation}
\Delta E\big|_{\Delta>1}
\lesssim
\sqrt{2}\,\|H\|
e^{-D_T/(2j_0)}.
\label{eq:gapped_bound}
\end{equation}

Therefore, the BA energy error decreases exponentially with the
retained subspace dimension $D_T$. This prediction is consistent with
the small errors observed at $\Delta=2.0$ in
Table~\ref{tab:subspace_error}, despite retaining only a tiny
fraction of the full Hilbert space.

\subsection{Quasi-long-range ordered regime ($-1<\Delta\le1$)}
\label{sec:qlro}

For $-1<\Delta\le1$, the XXZ chain is in a gapless
Tomonaga--Luttinger-liquid phase characterized by algebraically
decaying correlations.

Numerically, we find that the sorted computational-basis
probabilities are well described by a power-law form,

\begin{equation}
|c_j|^2
\simeq
\frac{\tilde c_0}{j^\gamma},
\label{eq:critical_coeff}
\end{equation}

where $\gamma$ is obtained from fitting the coefficient distribution.
The slower decay compared to Eq.~\eqref{eq:gapped_coeff} indicates a
less sparse ground state and consequently a slower convergence of the
BA method.

As shown in Appendix~\ref{app:qlro}, the retained weight satisfies

\begin{equation}
\alpha_{D_T}
=
\frac{
1-D_T^{\,1-\gamma}
}{
1-D_{\mathcal H}^{\,1-\gamma}
}.
\end{equation}

The resulting convergence behavior depends on the value of the
coefficient exponent $\gamma$.

For $\gamma>1$,

\begin{equation}
\left(
1-\sqrt{\alpha_{D_T}}
\right)^{1/2}
\simeq
\frac{1}{\sqrt2}
D_T^{-(\gamma-1)/2},
\end{equation}

which yields

\begin{equation}
\Delta E
\lesssim
\sqrt2\,\|H\|\,
D_T^{-(\gamma-1)/2}.
\label{eq:critical_bound}
\end{equation}

The marginal case $\gamma=1$ separates the two regimes. In this case,

\begin{equation}
\alpha_{D_T}
=
\frac{\ln D_T}
     {\ln D_{\mathcal H}},
\end{equation}

which gives

\begin{equation}
\Delta E
\lesssim
2\|H\|
\left[
1-
\sqrt{
\frac{\ln D_T}
     {\ln D_{\mathcal H}}
}
\right]^{1/2}.
\label{eq:critical_bound_gamma_eq_1}
\end{equation}

The convergence is therefore logarithmically slow in the retained
subspace dimension $D_T$.

For $\gamma<1$,

\begin{equation}
\alpha_{D_T}
\simeq
\left(
\frac{D_T}{D_{\mathcal H}}
\right)^{1-\gamma},
\end{equation}

leading to

\begin{equation}
\Delta E
\lesssim
2\|H\|
\left[
1-
\left(
\frac{D_T}{D_{\mathcal H}}
\right)^{(1-\gamma)/2}
\right]^{1/2}.
\label{eq:critical_bound_gamma_lt_1}
\end{equation}

Thus, irrespective of the precise value of $\gamma$, the convergence
in the critical regime is substantially slower than the exponential
behavior observed in the gapped N\'eel phase. The slower decay of the
sorted coefficient distribution implies that significantly larger
subspaces are required to achieve a given target accuracy, consistent
with the larger BA errors observed near the critical region in
Table~\ref{tab:subspace_error}.

\paragraph{XY / ferromagnetic boundary ($\Delta\to-1$).}

As $\Delta\to-1^{+}$, the fitted coefficient exponent approaches the
marginal value $\gamma\to1^{+}$. The ground state becomes increasingly
dominated by a small number of nearly polarized configurations,
causing $\alpha_{D_T}\to1$ rapidly even for modest values of $D_T$.
This explains the low errors ($0.2$--$0.5\%$) observed at
$\Delta=-1.0$ despite the system being formally within the
quasi-long-range ordered phase.

\subsection{Scaling summary}

Table~\ref{tab:scaling_summary} summarizes the scaling of the retained
weight and the corresponding BA energy-error bounds in the different
regimes considered in this work. The exponential convergence in the
gapped N\'eel phase reflects the strong sparsity of the ground-state
wavefunction, whereas the algebraic or logarithmic convergence in the
critical phase originates from the slower decay of the sorted
computational-basis coefficients. Consequently, significantly larger
subspaces are required near criticality to achieve the same target
accuracy.

\begin{table}[h!]
\centering
\caption{Scaling of the retained weight and corresponding BA
energy-error bounds for the XXZ chain. Here $D_T$ denotes the retained
subspace dimension, $D_{\mathcal H}$ the full Hilbert-space dimension,
and $\gamma$ the exponent obtained from fitting the sorted
computational-basis coefficients,
$|c_j|^2\propto j^{-\gamma}$.}
\label{tab:scaling_summary}
\renewcommand{\arraystretch}{1.3}

\resizebox{\linewidth}{!}{%
\begin{tabular}{|c|c|c|}
\hline
\textbf{Regime} &
\textbf{$1-\sqrt{\alpha_{D_T}}$} &
\textbf{$\Delta E$ bound} \\
\hline

Gapped ($\Delta>1$)
&
$\displaystyle e^{-D_T/j_0}$
&
$\displaystyle \sqrt{2}\,\|H\|\,e^{-D_T/(2j_0)}$
\\
\hline

Critical ($\gamma>1$)
&
$\displaystyle D_T^{\,1-\gamma}$
&
$\displaystyle \sqrt{2}\,\|H\|\,D_T^{-(\gamma-1)/2}$
\\
\hline

Critical ($\gamma=1$)
&
$\displaystyle
1-\sqrt{\frac{\ln D_T}{\ln D_{\mathcal H}}}
$
&
$\displaystyle
2\|H\|
\left[
1-\sqrt{
\frac{\ln D_T}
     {\ln D_{\mathcal H}}
}
\right]^{1/2}
$
\\
\hline

Critical ($\gamma<1$)
&
$\displaystyle
1-
\left(
\frac{D_T}{D_{\mathcal H}}
\right)^\frac{{1-\gamma}}{2}
$
&
$\displaystyle
2\|H\|
\left[
1-
\left(
\frac{D_T}{D_{\mathcal H}}
\right)^{(1-\gamma)/2}
\right]^{1/2}
$
\\
\hline

XY / ferromagnetic boundary ($\Delta\to-1$)
&
$\displaystyle \alpha_{D_T}\rightarrow 1$
&
$\displaystyle \Delta E\rightarrow 0$
\\
\hline

\end{tabular}}
\end{table}

%-----------------------

\section{Results}
In this section we present a detailed analysis of the performance of the proposed quantum-classical hybrid BA algorithm applied to a antiferromagnetic spin-$\frac{1}{2}$ XXZ Heisenberg Hamiltonian on a chain. The initial basis states used to initialize the BA algorithm for different values of the anisotropy parameter $\Delta$ are summarized in Table~\ref{tab:init_basis}.
%We start with only two initial basis states ($m_0=2$); $\ket{b^0_1}=\ket{\uparrow \downarrow  \uparrow \downarrow \uparrow \downarrow ..}$ and $\ket{b^0_2}=\ket{\downarrow \uparrow \downarrow \uparrow \downarrow ..}$ for $\Delta=1.0$ and $2.0$. For $\Delta=0.5$, we start with five initial states ($m_0=5$); $\ket{b^0_1}=\ket{\uparrow \downarrow  \uparrow \downarrow \uparrow \downarrow ..}$, $\ket{b^0_2}=\ket{\downarrow \uparrow \downarrow \uparrow \downarrow ..}$, $\ket{b^0_3}=\ket{\uparrow \uparrow \uparrow .. (N/2) \downarrow \downarrow \downarrow .. (N/2)}$, $\ket{b^0_4}=\ket{\downarrow \downarrow \downarrow .. (N/2) \uparrow \uparrow \uparrow .. (N/2)}$ and $\ket{b^0_5}=\frac{1}{\sqrt{2}}(\ket{\uparrow\downarrow}-\ket{\downarrow\uparrow})_{12}\otimes\frac{1}{\sqrt{2}}(\ket{\uparrow\downarrow}-\ket{\downarrow\uparrow})_{34}\otimes ...$. For $\Delta = -1$, we start with two initial states ($m_0 = 2$); $\ket{b^0_1}=\ket{\uparrow \uparrow \uparrow .. (N/2) \downarrow \downarrow \downarrow .. (N/2)}$, $\ket{b^0_2}=\ket{\downarrow \downarrow \downarrow .. (N/2) \uparrow \uparrow \uparrow .. (N/2)}$.

%--------------------
\begin{table}[h!]
\caption{Initial basis states used to initialize the BA algorithm for different values of the anisotropy parameter $\Delta$. Here, $m_0$ denotes the number of initial basis states.}
\label{tab:init_basis}
\centering
\small
\begin{tabular}{|c| c| l|}
\hline
$\Delta$ & $m_0$ & Initial basis states \\
\hline

$1.0,\;2.0$
&
2
&
\parbox[t]{0.68\columnwidth}{
$\ket{b_1^0}=\ket{\uparrow\downarrow\uparrow\downarrow\cdots}$,

$\ket{b_2^0}=\ket{\downarrow\uparrow\downarrow\uparrow\cdots}$.
}
\\[2mm]
\hline
$0.5$
&
5
&
\parbox[t]{0.68\columnwidth}{
$\ket{b_1^0}=\ket{\uparrow\downarrow\uparrow\downarrow\cdots}$,

$\ket{b_2^0}=\ket{\downarrow\uparrow\downarrow\uparrow\cdots}$,

$\ket{b_3^0}=\ket{\uparrow^{N/2}\downarrow^{N/2}}$,

$\ket{b_4^0}=\ket{\downarrow^{N/2}\uparrow^{N/2}}$,

$\ket{b_5^0}
=\displaystyle
\bigotimes_{i=1}^{N/2}
\frac{1}{\sqrt2}
\left(
\ket{\uparrow\downarrow}
-
\ket{\downarrow\uparrow}
\right)_{2i-1,\,2i}$.
}
\\[2mm]
\hline

$-1.0$
&
2
&
\parbox[t]{0.68\columnwidth}{
$\ket{b_1^0}=\ket{\uparrow^{N/2}\downarrow^{N/2}}$,

$\ket{b_2^0}=\ket{\downarrow^{N/2}\uparrow^{N/2}}$.
}
\\

\hline
\end{tabular}
\end{table}
%---------

The time evolution of these basis states or bitstrings are done using IBM quantum hardware (Heron) for system sizes up to $N=62$ qubits and we follow all the steps of the BA algorithm. We notice that even for the second iteration of the BA algorithm the error of the gs energy is reduced significantly. Therefore, all results are shown for the second iteration of algorithm and at the end of first iteration which start choosing only $m_1$ most probable basis obtained at the end of first iteration. We benchmark hybrid algorithm results against the density matrix renormalization group (DMRG) to evaluate the accuracy of ground state energies, and the reliability of spin-spin correlation functions.  We also compare  the energies with the Sampling Krylov Quantum Diagonalization (SKQD) method.

%----------

\subsection{Analysis of the gs energy}
%\subsubsection{Time evolution and symmetry-preserving sampling}
To generate the time-evolved state, we implemented a single-step first-order Trotterized evolution with a time step $\Delta t = 0.15$ for $N=48$ and $\Delta t = 0.12$ for $N=62$ [see Fig.\ref{fig:trotter_xxz}]. The evolved state was repeatedly measured in the computational basis, and the collected bitstrings were post-selected to enforce two symmetries inherent to the XXZ Hamiltonian: first, conservation of the z-component of the total spin $S^{z}$ and second, reflection symmetry of the spin chain. Enforcing symmetry at the sampling stage significantly stabilizes the reconstruction of the reduced Hilbert space and suppresses noise-induced leakage into symmetry-forbidden sectors. \\ 

%\subsubsection{Energy accuracy and fidelity with reduced basis dimension}
For all the calculation presented in this work is done by constructing the Hamiltonian matrix in reduced dimension of Hilbert space at the end of second iteration of the BA algorithm. Then the ground state energies and wavefunctions are calculated. In Fig. \ref{fig:percentage_error_states} (b) the percentage error in the gs energy is shown as a function of the number of the percentage of the total Hilbert space dimension. The percentage error in gs energy is defined as, $\Delta E_{gs}=\frac{E_{gs}(BA)-E_{gs}(DMRG)}{E_{gs}(DMRG)} \times 100$ where $E_{gs}(BA)$ and $E_{gs}(DMRG)$ are total gs energy calculated from the BA algorithm and  DMRG method respectively. $\Delta E_{gs}$ decreases rapidly as the number of basis states increases.
%\textcolor{red}{For $m_1=96$ and $\Delta E_{gs}$ is reduced to approximately $1\%$ and this accuracy is achieved only by recovering around $18\%$  of the total Hilbert space after $2^{nd}$ iteration shown in Table.}\ref{tab:subspace_error}. \\

There are two factors that affect the generation of new basis set, first the number of initial basis set at the beginning of $(i+1)^{th}$ iteration $m_i$ and the number of measurement shots $M_s$. The larger number of $m_i$ helps to generate the larger amount of basis states faster, whereas the higher number of shots will generate new basis states, but these states may include spin flips of higher order from the original basis state. 
%\textcolor{red}{In Fig. \ref{fig:percentage_error_states} (a), we show the dependence of $\Delta E_{gs}$ on $m_1$ for two $M_s=$ 20K and 40K and $\Delta=1.0$. Increasing $m_1$ expands dimension of relevant reduced Hilbert space at the second step and resulting higher accuracy of energy as shown in Fig. \ref{fig:percentage_error_states} (a). The higher number of $M_s$ lead to reducing sampling fluctuations, highlighting the importance of measurement statistics in the reconstruction process.}

%\textcolor{red}{We also evaluate the gs fidelity $F$ which is used to characterize the wavefunction accuracy and it can be defined as}

%\begin{equation}
%    F = \langle \psi_{gs}^{\mathrm{BAA}} | \psi_{gs}^{\mathrm{ED}} \rangle,
%\end{equation}

%\textcolor{red}{where $\ket{\psi_{gs}^{\mathrm{BAA}}}$ and $\ket{\psi_{gs}^{\mathrm{ED}}}$ are the gs wavefunctions obtained from BA algorithm and ED, respectively. The variation of fidelity with $\Delta$ is summarized in Table~\ref{tab:subspace_error}. For all the cases the Fidelity is always more the 0.96 which indicates excellent accuracy of the wavefunction.} 

\begin{table}[h!]
\centering
\caption{Subspace dimensions, referring to the basis generated after the second iteration, and percentage error in the ground-state energy ($\Delta E_{gs}$) for different system sizes ($N=48$, $62$)  and anisotropy values $\Delta$ with $M_s=60k$ and $m_1=96$.}
\label{tab:subspace_error}

% Increase spacing for a more expanded table
\renewcommand{\arraystretch}{1.35}
\setlength{\tabcolsep}{12pt}

\begin{tabular}{|c|cc|cc|}
\hline
& \multicolumn{2}{c|}{$N=48$} & \multicolumn{2}{c|}{$N=62$} \\
\cline{2-5}
$\Delta$ & \textbf{SD} & \textbf{$\Delta E_{gs}$} 
         & \textbf{SD} & \textbf{$\Delta E_{gs}$} \\
\hline
-1.0 & 160326 & 0.2\% & 201454 & 0.5\% \\
0.5 & 3469058 & 20.6\% & 3274338 & 28.7\% \\
1.0 & 3317688 & 10.7\% & 3182796 & 15.6\% \\
2.0 & 3913142 & 1.7\% & 3329260 & 3.5\% \\
\hline
\end{tabular}
\end{table}

\subsection{Dependence on anisotropy parameter $\Delta$}
\begin{figure}[h]
\includegraphics[width=1.0\columnwidth]{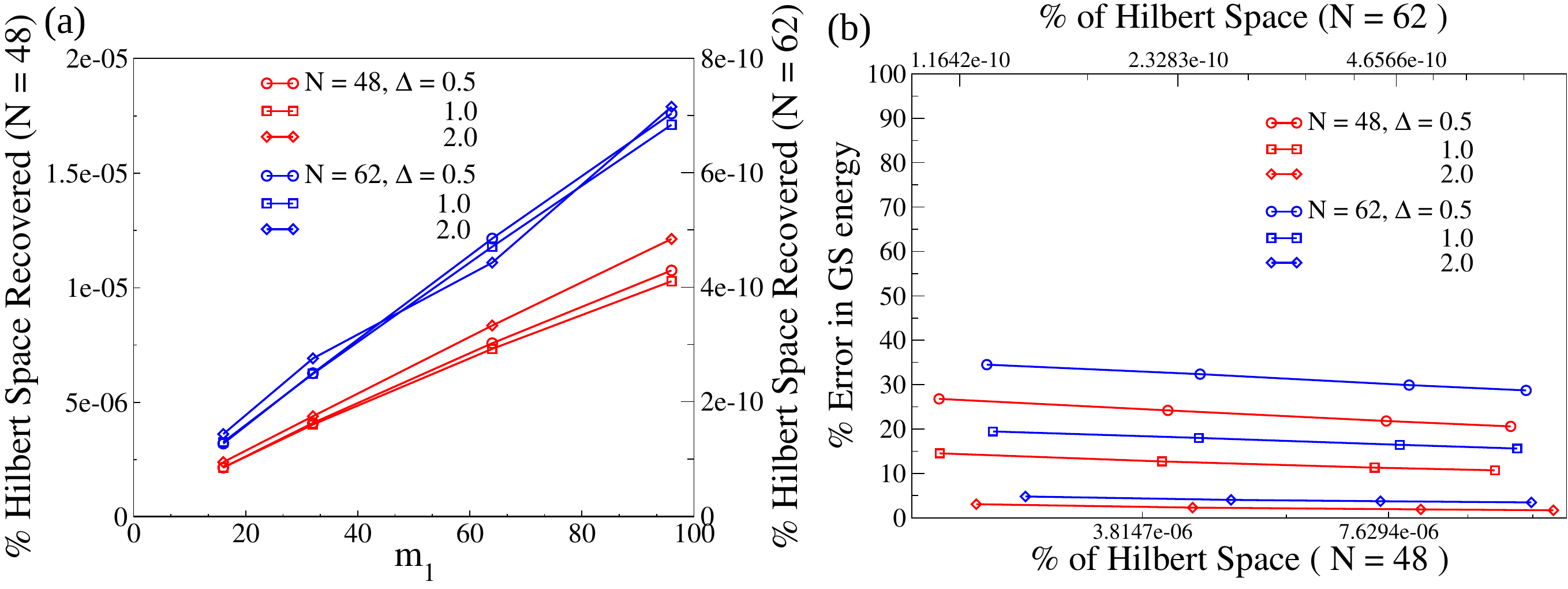}
\caption{(a) Percentage recovery of the Hilbert Space with the maximum number of retained basis states ($m_1$), shown for different values of $\Delta$ and $N=48$, $N=62$. (b) Percentage error in GS Energy as a function of $m_1$ for several values of $\Delta$, computed using $M_s = 60k$ shots. The results highlight how enlarging the basis enhances the ground-state accuracy.}

\label{fig:percentage_error_states}
\end{figure}

To examine the performance of the algorithm for various parameter regimes, we computed the ground-state energy for several values of the anisotropy parameter $\Delta$. Fig. \ref{fig:percentage_error_states} (a) displays the percentage recovery of the total Hilbert space for different values of $m_1$ ranging from $16$ to $96$, with $M_s=60k$. On the other hand, Fig. \ref{fig:percentage_error_states} (b) displays the percentage deviation of gs energy from the DMRG energy for $\Delta = -1.0, 0.5, 1.0, 2.0$, as a function of the percentage of the total Hilbert space dimension, with $M_s=60k$ shots.  

For the antiferromagnetic isotropic Heisenberg limit, $\Delta = 1.0$, strong quantum fluctuations lead to a broad distribution of basis states, and the ground state energy error is around $10\%$ for $m_1=96$ and $N=48$, while it increases to $20\%$ for $N=62$ as shown in Fig. \ref{fig:percentage_error_states} (b). Increasing the values of $\Delta$ the system gradually enters the Ising-dominated regime, where the longitudinal interaction term $\Delta S_{i}^{z} S_{i+1}^{z}$ dominates and the transverse fluctuation decreases. In this regime, the ground state becomes increasingly classical and spin configuration is close to antiferromagnet. Consequently, the sampling-based reconstruction becomes more efficient, and the error in the ground-state energy systematically decreases. We also notice that $m_1=64$ is sufficient to get the energy error less than 2\% for $N=48$.

\subsection{Spin-Spin Correlation Functions}
Generally, error in global quantities like total energy tend to cancel out and give very good accuracy in the approximate method but local correlation accuracy are generally poor. Therefore we calculate the spin-spin correlation function correlation function $C(r)$ between two spins separated by a distance $r$, which provides valuable insight into the nature of magnetic ordering in the system. The correlation function is defined as
\begin{equation}
    C(r) = \langle \vec{S}_i \cdot \vec{S}_{i+r} \rangle.
\end{equation}

\begin{figure}[h]
\includegraphics[width=1.0\columnwidth]{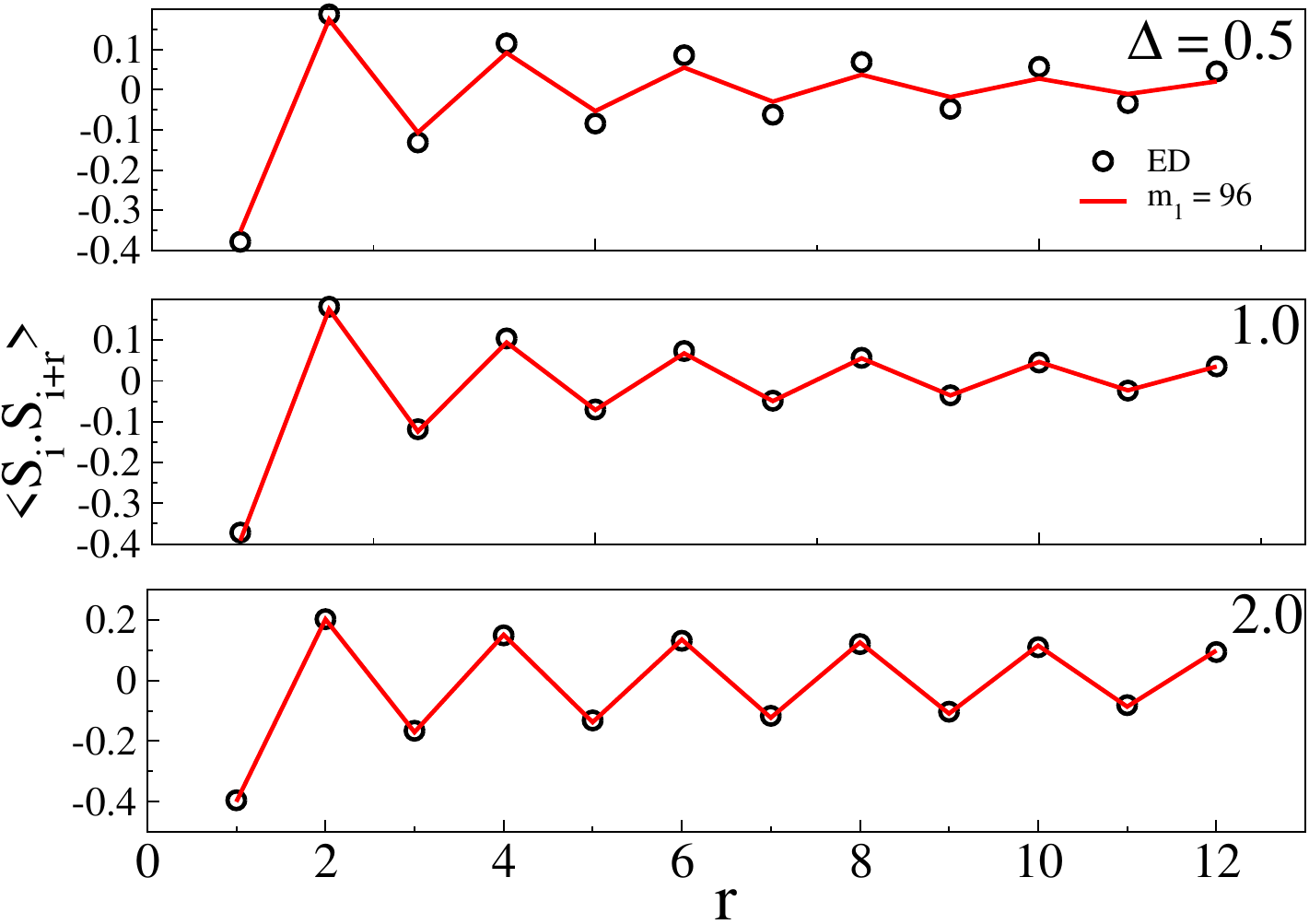}
\caption{For different $\Delta$ (= 0.5, 1.0 and 2.0) values correlation function is shown for 24 number of qubits with $M_s=40k$.
}
\label{fig:corsz_comp}
\end{figure}

Fig.~\ref{fig:corsz_comp} shows $C(r)$ for a system size $N = 24$ and several values of the anisotropy parameter $\Delta$, computed using $M_s=40k$ and $m_1 = 96$. In Fig.~\ref{fig:corsz_comp}, the results obtained from ED (shown as circles) are compared with those from BA algorithm (shown as lines). For all values of $\Delta$, the correlation functions exhibit excellent agreement with the ED benchmarks. The reliable reproduction of $C(r)$ over long distances indicates that the algorithm not only captures global energetic properties but also accurately reconstructs non-local correlations, which are sensitive probes of the wavefunction structure.
% \begin{table}[h!]
% \centering
% \caption{Fidelity of the reconstructed ground state for $N=24$ with 40K shots and $m=96$.}
% \label{tab:fidelity_only}
% \footnotesize
% \begin{tabular}{c|c}
% \hline
% $\Delta$ & \textbf{Fidelity} \\
% \hline
% 1.0 & 0.96534 \\
% 1.2 & 0.97404 \\
% 1.4 & 0.98534 \\
% \hline
% \end{tabular}
% \end{table}

\subsection{Comparison with SKQD}
To benchmark the efficiency and accuracy of the BA algorithm, we compare its performance with the Sample-Based Krylov Quantum Diagonalization (SKQD) method, one of the leading quantum-assisted subspace diagonalization approaches currently available \cite{misciasci2026}. We focus on the most challenging regime of the XXZ chain, namely the isotropic point ($\Delta=1$) at zero longitudinal field ($h_z=0$), where the ground state is highly entangled and exhibits the weakest computational-basis sparsity.

The SKQD benchmark employs a Krylov subspace of dimension $d=5$ generated using real-time evolution with time step $\Delta t=0.06$. Each Krylov vector is implemented using a second-order Trotter decomposition with three Trotter repetitions. For the largest one-dimensional benchmark reported in Ref.~\cite{misciasci2026}, $60k$ shots were used per Krylov vector, corresponding to a total sampling cost of approximately $300k$ shots. Despite this substantial sampling effort, the relative ground-state energy error remains significant in the low-field isotropic regime.

Table~\ref{tab:skqd_vs_ours} compares the ground-state energy errors obtained by SKQD and the BA algorithm. For $N=24$, the SKQD error extracted from Fig.~5 of Ref.~\cite{misciasci2026} is approximately $2.5\%$, whereas the BA algorithm achieves an error of only $0.87\%$. For larger systems, SKQD exhibits an error of approximately $35\%$ for $N=50$, while the BA algorithm yields a substantially smaller error of $10.70\%$ for the closely comparable system size $N=48$.

The superior performance of the BA algorithm originates from its adaptive construction of a symmetry-preserving reduced Hilbert space. Unlike SKQD, which generates a Krylov basis through repeated time evolution and relies on extensive sampling to recover the relevant configurations, the BA algorithm directly identifies the most probable basis states generated during time evolution and enforces both $U(1)$ and lattice-reflection symmetries at every iteration. Consequently, the reduced Hamiltonian is constructed within a compact and physically relevant subspace, leading to a more stable classical diagonalization procedure and improved accuracy.

\begin{table}[h]
\centering
\label{tab:skqd_vs_ours}
\caption{Comparison between SKQD and the BA algorithm at the isotropic point
($\Delta=1$, $h_z=0$). The SKQD errors are estimated from Fig.~5 of
Ref.~\cite{misciasci2026}.}
\label{tab:skqd_vs_ours}
\begin{tabular}{|c|c|c|}
\hline
System size & SKQD error & BA error \\
\hline
$N=24$ & $\sim 2.5\%$ & $0.87\%$ \\
$N=48$ ($N=50$ for SKQD) & $\sim 35\%$ & $10.70\%$ \\
\hline
\end{tabular}
\end{table}

Furthermore, the BA algorithm employs significantly shallower circuits than SKQD. Time evolution is used solely to generate relevant configurations rather than to construct and sample multiple Krylov vectors. This substantially reduces circuit depth and measurement overhead while retaining the physically important low-energy sector. 

\section{Summary \& Future Work}
The challenge of simulating correlated quantum many-body systems stems from the exponential scaling of the Hilbert space. While quantum computation offers a path forward, existing algorithms like VQE are limited by complex optimization landscapes, and QPE requires prohibitive circuit depth. We introduce an hybrid quantum-classical algorithm ,the BA algorithm' which overcomes above limitations by  utilizing shallow Trotterized circuits for short and one step real-time evolution on a quantum processor and choosing the most relevant basis for the selected energy state. The resulting relevant basis of the reduced Hilbert space are then rigorously symmetry-filtered to enforce symmetry like conservation of longitudinal component of total spin  and lattice reflection symmetry of the model Hamiltonian. Thereafter, classical computer diagonalizes the Hamiltonian within reduced Hilbert space to find the gs properties.

Benchmarking of the gs results for the spin-1/2 XXZ chain up to $N=62$ qubits (on IBM's Heron) demonstrates the algorithm achieves good accuracy in ground-state energies across anisotropy values $-1\le \Delta \le 2$.  At the isotropic Heisenberg point, the proposed algorithm achieves an error of $10.70\%$ for $N=48$, while the SKQD algorithm reports a substantially larger error of approximately $35\%$ for $N=50$ \cite{misciasci2026}. For $N=62$ the BAA algorithm yields errors of $0.5\%$ and $15.6\%$ at $\Delta=-1.0$ and $\Delta=1.0$, respectively. The algorithm exhibits the maximum error of 28.7\% at $\Delta=0.5$ (see Table~\ref{tab:subspace_error}), owing to the quasi-long-range nature of the system in this parameter regime. In this case, the chosen subspace dimension is insufficient to fully capture all the correlations. The method also accurately reproduces spin-spin correlation functions and outperforms Sampling Krylov Quantum Diagonalization (SKQD). This validates that symmetry-filtered, real-time quantum sampling offers a robust and efficient pathway for high-accuracy ground-state simulations on current near-term hardware. 

 The BA algorithm can be extended to more challenging quantum many-body systems, including two-dimensional frustrated spin lattices, strongly correlated fermionic models such as the Hubbard model, and \textit{ab initio} molecular Hamiltonians. Its adaptive, symmetry-filtered basis construction provides a scalable framework for investigating larger quantum systems on near-term quantum hardware and can be further generalized to study excited states and quantum dynamics.

\section{Acknowledgment}
We thank Dr. Richa Goel, Dr. Sieglinde Pfaendler, Dr. Shesha Raghunathan for their fruitful discussions. AB acknowledges the financial support from DST-India. SG thanks DST-Inspire for financial support. We acknowledge the National Supercomputing Mission (NSM) for providing computing resources of ‘PARAM RUDRA’ at S.N. Bose National Centre for Basic Sciences, which is implemented by C-DAC and supported by the Ministry of Electronics and Information Technology (MeitY) and the Department of Science and Technology (DST), Government of India. We acknowledge the use of IBM Quantum Credits for this work.
\bibliography{ref}

\begin{thebibliography}{59}%
\makeatletter
\providecommand \@ifxundefined [1]{%
 \@ifx{#1\undefined}
}%
\providecommand \@ifnum [1]{%
 \ifnum #1\expandafter \@firstoftwo
 \else \expandafter \@secondoftwo
 \fi
}%
\providecommand \@ifx [1]{%
 \ifx #1\expandafter \@firstoftwo
 \else \expandafter \@secondoftwo
 \fi
}%
\providecommand \natexlab [1]{#1}%
\providecommand \enquote  [1]{``#1''}%
\providecommand \bibnamefont  [1]{#1}%
\providecommand \bibfnamefont [1]{#1}%
\providecommand \citenamefont [1]{#1}%
\providecommand \href@noop [0]{\@secondoftwo}%
\providecommand \href [0]{\begingroup \@sanitize@url \@href}%
\providecommand \@href[1]{\@@startlink{#1}\@@href}%
\providecommand \@@href[1]{\endgroup#1\@@endlink}%
\providecommand \@sanitize@url [0]{\catcode `\\12\catcode `\$12\catcode
  `\&12\catcode `\#12\catcode `\^12\catcode `\_12\catcode `\%12\relax}%
\providecommand \@@startlink[1]{}%
\providecommand \@@endlink[0]{}%
\providecommand \url  [0]{\begingroup\@sanitize@url \@url }%
\providecommand \@url [1]{\endgroup\@href {#1}{\urlprefix }}%
\providecommand \urlprefix  [0]{URL }%
\providecommand \Eprint [0]{\href }%
\providecommand \doibase [0]{https://doi.org/}%
\providecommand \selectlanguage [0]{\@gobble}%
\providecommand \bibinfo  [0]{\@secondoftwo}%
\providecommand \bibfield  [0]{\@secondoftwo}%
\providecommand \translation [1]{[#1]}%
\providecommand \BibitemOpen [0]{}%
\providecommand \bibitemStop [0]{}%
\providecommand \bibitemNoStop [0]{.\EOS\space}%
\providecommand \EOS [0]{\spacefactor3000\relax}%
\providecommand \BibitemShut  [1]{\csname bibitem#1\endcsname}%
\let\auto@bib@innerbib\@empty
%</preamble>
\bibitem [{\citenamefont {Balents}(2010)}]{balents2010spinliquids}%
  \BibitemOpen
  \bibfield  {author} {\bibinfo {author} {\bibfnamefont {L.}~\bibnamefont
  {Balents}},\ }\bibfield  {title} {\bibinfo {title} {Spin liquids in
  frustrated magnets},\ }\href@noop {} {\bibfield  {journal} {\bibinfo
  {journal} {Nature}\ }\textbf {\bibinfo {volume} {464}},\ \bibinfo {pages}
  {199} (\bibinfo {year} {2010})}\BibitemShut {NoStop}%
\bibitem [{\citenamefont {Moessner}\ and\ \citenamefont
  {Ramirez}(2006)}]{moessner2006frustration}%
  \BibitemOpen
  \bibfield  {author} {\bibinfo {author} {\bibfnamefont {R.}~\bibnamefont
  {Moessner}}\ and\ \bibinfo {author} {\bibfnamefont {A.~P.}\ \bibnamefont
  {Ramirez}},\ }\bibfield  {title} {\bibinfo {title} {Geometrical
  frustration},\ }\href@noop {} {\bibfield  {journal} {\bibinfo  {journal}
  {Physics Today}\ }\textbf {\bibinfo {volume} {59}},\ \bibinfo {pages} {24}
  (\bibinfo {year} {2006})}\BibitemShut {NoStop}%
\bibitem [{\citenamefont {Gardner}\ \emph {et~al.}(2010)\citenamefont
  {Gardner}, \citenamefont {Gingras},\ and\ \citenamefont
  {Greedan}}]{gardner2010pyrochlore}%
  \BibitemOpen
  \bibfield  {author} {\bibinfo {author} {\bibfnamefont {J.~S.}\ \bibnamefont
  {Gardner}}, \bibinfo {author} {\bibfnamefont {M.~J.}\ \bibnamefont
  {Gingras}},\ and\ \bibinfo {author} {\bibfnamefont {J.~E.}\ \bibnamefont
  {Greedan}},\ }\bibfield  {title} {\bibinfo {title} {Magnetic pyrochlore
  oxides},\ }\href@noop {} {\bibfield  {journal} {\bibinfo  {journal} {Rev.
  Mod. Phys.}\ }\textbf {\bibinfo {volume} {82}},\ \bibinfo {pages} {53}
  (\bibinfo {year} {2010})}\BibitemShut {NoStop}%
\bibitem [{\citenamefont {Lee}\ \emph {et~al.}(2006)\citenamefont {Lee},
  \citenamefont {Nagaosa},\ and\ \citenamefont {Wen}}]{lee2006htsc}%
  \BibitemOpen
  \bibfield  {author} {\bibinfo {author} {\bibfnamefont {P.~A.}\ \bibnamefont
  {Lee}}, \bibinfo {author} {\bibfnamefont {N.}~\bibnamefont {Nagaosa}},\ and\
  \bibinfo {author} {\bibfnamefont {X.-G.}\ \bibnamefont {Wen}},\ }\bibfield
  {title} {\bibinfo {title} {Doping a mott insulator: Physics of
  high-temperature superconductivity},\ }\href@noop {} {\bibfield  {journal}
  {\bibinfo  {journal} {Rev. Mod. Phys.}\ }\textbf {\bibinfo {volume} {78}},\
  \bibinfo {pages} {17} (\bibinfo {year} {2006})}\BibitemShut {NoStop}%
\bibitem [{\citenamefont {Scalapino}(2012)}]{scalapino2012unconventional}%
  \BibitemOpen
  \bibfield  {author} {\bibinfo {author} {\bibfnamefont {D.~J.}\ \bibnamefont
  {Scalapino}},\ }\bibfield  {title} {\bibinfo {title} {A common thread: The
  pairing interaction for unconventional superconductors},\ }\href@noop {}
  {\bibfield  {journal} {\bibinfo  {journal} {Rev. Mod. Phys.}\ }\textbf
  {\bibinfo {volume} {84}},\ \bibinfo {pages} {1383} (\bibinfo {year}
  {2012})}\BibitemShut {NoStop}%
\bibitem [{\citenamefont {Keimer}\ \emph {et~al.}(2015)\citenamefont {Keimer},
  \citenamefont {Kivelson}, \citenamefont {Norman}, \citenamefont {Uchida},\
  and\ \citenamefont {Zaanen}}]{keimer2015review}%
  \BibitemOpen
  \bibfield  {author} {\bibinfo {author} {\bibfnamefont {B.}~\bibnamefont
  {Keimer}}, \bibinfo {author} {\bibfnamefont {S.~A.}\ \bibnamefont
  {Kivelson}}, \bibinfo {author} {\bibfnamefont {M.~R.}\ \bibnamefont
  {Norman}}, \bibinfo {author} {\bibfnamefont {S.-i.}\ \bibnamefont {Uchida}},\
  and\ \bibinfo {author} {\bibfnamefont {J.}~\bibnamefont {Zaanen}},\
  }\bibfield  {title} {\bibinfo {title} {From quantum matter to
  high-temperature superconductivity in copper oxides},\ }\href@noop {}
  {\bibfield  {journal} {\bibinfo  {journal} {Nature}\ }\textbf {\bibinfo
  {volume} {518}},\ \bibinfo {pages} {179} (\bibinfo {year}
  {2015})}\BibitemShut {NoStop}%
\bibitem [{\citenamefont {Savary}\ and\ \citenamefont
  {Balents}(2017)}]{savary2017qsl}%
  \BibitemOpen
  \bibfield  {author} {\bibinfo {author} {\bibfnamefont {L.}~\bibnamefont
  {Savary}}\ and\ \bibinfo {author} {\bibfnamefont {L.}~\bibnamefont
  {Balents}},\ }\bibfield  {title} {\bibinfo {title} {Quantum spin liquids: a
  review},\ }\href@noop {} {\bibfield  {journal} {\bibinfo  {journal} {Rep.
  Prog. Phys.}\ }\textbf {\bibinfo {volume} {80}},\ \bibinfo {pages} {016502}
  (\bibinfo {year} {2017})}\BibitemShut {NoStop}%
\bibitem [{\citenamefont {Zhou}\ \emph {et~al.}(2017)\citenamefont {Zhou},
  \citenamefont {Kanoda},\ and\ \citenamefont {Ng}}]{zhou2017qsl}%
  \BibitemOpen
  \bibfield  {author} {\bibinfo {author} {\bibfnamefont {Y.}~\bibnamefont
  {Zhou}}, \bibinfo {author} {\bibfnamefont {K.}~\bibnamefont {Kanoda}},\ and\
  \bibinfo {author} {\bibfnamefont {T.-K.}\ \bibnamefont {Ng}},\ }\bibfield
  {title} {\bibinfo {title} {Quantum spin liquid states},\ }\href@noop {}
  {\bibfield  {journal} {\bibinfo  {journal} {Rev. Mod. Phys.}\ }\textbf
  {\bibinfo {volume} {89}},\ \bibinfo {pages} {025003} (\bibinfo {year}
  {2017})}\BibitemShut {NoStop}%
\bibitem [{\citenamefont {Yan}\ \emph {et~al.}(2011)\citenamefont {Yan},
  \citenamefont {Huse},\ and\ \citenamefont {White}}]{yan2011kagome}%
  \BibitemOpen
  \bibfield  {author} {\bibinfo {author} {\bibfnamefont {S.}~\bibnamefont
  {Yan}}, \bibinfo {author} {\bibfnamefont {D.~A.}\ \bibnamefont {Huse}},\ and\
  \bibinfo {author} {\bibfnamefont {S.~R.}\ \bibnamefont {White}},\ }\bibfield
  {title} {\bibinfo {title} {Spin-liquid ground state of the s=1/2 kagome
  heisenberg antiferromagnet},\ }\href@noop {} {\bibfield  {journal} {\bibinfo
  {journal} {Science}\ }\textbf {\bibinfo {volume} {332}},\ \bibinfo {pages}
  {1173} (\bibinfo {year} {2011})}\BibitemShut {NoStop}%
\bibitem [{\citenamefont {Anderson}(1987)}]{anderson1987resonating}%
  \BibitemOpen
  \bibfield  {author} {\bibinfo {author} {\bibfnamefont {P.~W.}\ \bibnamefont
  {Anderson}},\ }\bibfield  {title} {\bibinfo {title} {The resonating valence
  bond state in la2cuo4 and superconductivity},\ }\href@noop {} {\bibfield
  {journal} {\bibinfo  {journal} {science}\ }\textbf {\bibinfo {volume}
  {235}},\ \bibinfo {pages} {1196} (\bibinfo {year} {1987})}\BibitemShut
  {NoStop}%
\bibitem [{\citenamefont {Imada}\ \emph {et~al.}(1998)\citenamefont {Imada},
  \citenamefont {Fujimori},\ and\ \citenamefont {Tokura}}]{ImadaRMP}%
  \BibitemOpen
  \bibfield  {author} {\bibinfo {author} {\bibfnamefont {M.}~\bibnamefont
  {Imada}}, \bibinfo {author} {\bibfnamefont {A.}~\bibnamefont {Fujimori}},\
  and\ \bibinfo {author} {\bibfnamefont {Y.}~\bibnamefont {Tokura}},\
  }\bibfield  {title} {\bibinfo {title} {Metal-insulator transitions},\
  }\href@noop {} {\bibfield  {journal} {\bibinfo  {journal} {Rev. Mod. Phys.}\
  }\textbf {\bibinfo {volume} {70}},\ \bibinfo {pages} {1039} (\bibinfo {year}
  {1998})}\BibitemShut {NoStop}%
\bibitem [{\citenamefont {Mahan}(2000)}]{mah00}%
  \BibitemOpen
  \bibfield  {author} {\bibinfo {author} {\bibfnamefont {G.~D.}\ \bibnamefont
  {Mahan}},\ }\href@noop {} {\emph {\bibinfo {title} {Many Particle Physics,
  Third Edition}}}\ (\bibinfo  {publisher} {Plenum},\ \bibinfo {address} {New
  York},\ \bibinfo {year} {2000})\BibitemShut {NoStop}%
\bibitem [{\citenamefont {Fetter}\ and\ \citenamefont
  {Walecka}(1971)}]{Fetter}%
  \BibitemOpen
  \bibfield  {author} {\bibinfo {author} {\bibfnamefont {A.~L.}\ \bibnamefont
  {Fetter}}\ and\ \bibinfo {author} {\bibfnamefont {J.~D.}\ \bibnamefont
  {Walecka}},\ }\href@noop {} {\emph {\bibinfo {title} {Quantum Theory of
  Many-Particle Systems}}}\ (\bibinfo  {publisher} {McGraw-Hill},\ \bibinfo
  {address} {Boston},\ \bibinfo {year} {1971})\BibitemShut {NoStop}%
\bibitem [{\citenamefont
  {Sandvik}(2010{\natexlab{a}})}]{sandvik2010computational}%
  \BibitemOpen
  \bibfield  {author} {\bibinfo {author} {\bibfnamefont {A.~W.}\ \bibnamefont
  {Sandvik}},\ }\bibfield  {title} {\bibinfo {title} {Computational studies of
  quantum spin systems},\ }in\ \href@noop {} {\emph {\bibinfo {booktitle} {AIP
  Conference Proceedings}}},\ Vol.\ \bibinfo {volume} {1297}\ (\bibinfo
  {organization} {American Institute of Physics},\ \bibinfo {year} {2010})\
  pp.\ \bibinfo {pages} {135--338}\BibitemShut {NoStop}%
\bibitem [{\citenamefont {Lin}(1990)}]{lin1990exact}%
  \BibitemOpen
  \bibfield  {author} {\bibinfo {author} {\bibfnamefont {H.}~\bibnamefont
  {Lin}},\ }\bibfield  {title} {\bibinfo {title} {Exact diagonalization of
  quantum-spin models},\ }\href@noop {} {\bibfield  {journal} {\bibinfo
  {journal} {Physical Review B}\ }\textbf {\bibinfo {volume} {42}},\ \bibinfo
  {pages} {6561} (\bibinfo {year} {1990})}\BibitemShut {NoStop}%
\bibitem [{\citenamefont {Nataf}\ and\ \citenamefont
  {Mila}(2014)}]{nataf2014exact}%
  \BibitemOpen
  \bibfield  {author} {\bibinfo {author} {\bibfnamefont {P.}~\bibnamefont
  {Nataf}}\ and\ \bibinfo {author} {\bibfnamefont {F.}~\bibnamefont {Mila}},\
  }\bibfield  {title} {\bibinfo {title} {Exact diagonalization of heisenberg su
  (n) models},\ }\href@noop {} {\bibfield  {journal} {\bibinfo  {journal}
  {Physical review letters}\ }\textbf {\bibinfo {volume} {113}},\ \bibinfo
  {pages} {127204} (\bibinfo {year} {2014})}\BibitemShut {NoStop}%
\bibitem [{\citenamefont {Foulkes}\ \emph {et~al.}(2001)\citenamefont
  {Foulkes}, \citenamefont {Mitas}, \citenamefont {Needs},\ and\ \citenamefont
  {Rajagopal}}]{foulkes2001quantum}%
  \BibitemOpen
  \bibfield  {author} {\bibinfo {author} {\bibfnamefont {W.~M.}\ \bibnamefont
  {Foulkes}}, \bibinfo {author} {\bibfnamefont {L.}~\bibnamefont {Mitas}},
  \bibinfo {author} {\bibfnamefont {R.}~\bibnamefont {Needs}},\ and\ \bibinfo
  {author} {\bibfnamefont {G.}~\bibnamefont {Rajagopal}},\ }\bibfield  {title}
  {\bibinfo {title} {Quantum monte carlo simulations of solids},\ }\href@noop
  {} {\bibfield  {journal} {\bibinfo  {journal} {Reviews of Modern Physics}\
  }\textbf {\bibinfo {volume} {73}},\ \bibinfo {pages} {33} (\bibinfo {year}
  {2001})}\BibitemShut {NoStop}%
\bibitem [{\citenamefont {Pollet}(2012)}]{pollet2012recent}%
  \BibitemOpen
  \bibfield  {author} {\bibinfo {author} {\bibfnamefont {L.}~\bibnamefont
  {Pollet}},\ }\bibfield  {title} {\bibinfo {title} {Recent developments in
  quantum monte carlo simulations with applications for cold gases},\
  }\href@noop {} {\bibfield  {journal} {\bibinfo  {journal} {Reports on
  progress in physics}\ }\textbf {\bibinfo {volume} {75}},\ \bibinfo {pages}
  {094501} (\bibinfo {year} {2012})}\BibitemShut {NoStop}%
\bibitem [{\citenamefont {Gubernatis}\ \emph {et~al.}(2016)\citenamefont
  {Gubernatis}, \citenamefont {Kawashima},\ and\ \citenamefont
  {Werner}}]{gubernatis2016quantum}%
  \BibitemOpen
  \bibfield  {author} {\bibinfo {author} {\bibfnamefont {J.}~\bibnamefont
  {Gubernatis}}, \bibinfo {author} {\bibfnamefont {N.}~\bibnamefont
  {Kawashima}},\ and\ \bibinfo {author} {\bibfnamefont {P.}~\bibnamefont
  {Werner}},\ }\href@noop {} {\emph {\bibinfo {title} {Quantum Monte Carlo
  Methods}}}\ (\bibinfo  {publisher} {Cambridge University Press},\ \bibinfo
  {year} {2016})\BibitemShut {NoStop}%
\bibitem [{\citenamefont {Sandvik}\ and\ \citenamefont
  {Kurkij{\"a}rvi}(1991)}]{sandvik1991quantum}%
  \BibitemOpen
  \bibfield  {author} {\bibinfo {author} {\bibfnamefont {A.~W.}\ \bibnamefont
  {Sandvik}}\ and\ \bibinfo {author} {\bibfnamefont {J.}~\bibnamefont
  {Kurkij{\"a}rvi}},\ }\bibfield  {title} {\bibinfo {title} {Quantum monte
  carlo simulation method for spin systems},\ }\href@noop {} {\bibfield
  {journal} {\bibinfo  {journal} {Physical Review B}\ }\textbf {\bibinfo
  {volume} {43}},\ \bibinfo {pages} {5950} (\bibinfo {year}
  {1991})}\BibitemShut {NoStop}%
\bibitem [{\citenamefont {Ceperley}\ and\ \citenamefont
  {Alder}(1986)}]{ceperley1986quantum}%
  \BibitemOpen
  \bibfield  {author} {\bibinfo {author} {\bibfnamefont {D.}~\bibnamefont
  {Ceperley}}\ and\ \bibinfo {author} {\bibfnamefont {B.}~\bibnamefont
  {Alder}},\ }\bibfield  {title} {\bibinfo {title} {Quantum monte carlo},\
  }\href@noop {} {\bibfield  {journal} {\bibinfo  {journal} {Science}\ }\textbf
  {\bibinfo {volume} {231}},\ \bibinfo {pages} {555} (\bibinfo {year}
  {1986})}\BibitemShut {NoStop}%
\bibitem [{\citenamefont {Santos}(2003)}]{santos2003introduction}%
  \BibitemOpen
  \bibfield  {author} {\bibinfo {author} {\bibfnamefont {R.~R.~d.}\
  \bibnamefont {Santos}},\ }\bibfield  {title} {\bibinfo {title} {Introduction
  to quantum monte carlo simulations for fermionic systems},\ }\href@noop {}
  {\bibfield  {journal} {\bibinfo  {journal} {Brazilian Journal of Physics}\
  }\textbf {\bibinfo {volume} {33}},\ \bibinfo {pages} {36} (\bibinfo {year}
  {2003})}\BibitemShut {NoStop}%
\bibitem [{\citenamefont {Montangero}\ \emph {et~al.}(2018)\citenamefont
  {Montangero}, \citenamefont {Montangero},\ and\ \citenamefont
  {Evenson}}]{montangero2018introduction}%
  \BibitemOpen
  \bibfield  {author} {\bibinfo {author} {\bibfnamefont {S.}~\bibnamefont
  {Montangero}}, \bibinfo {author} {\bibfnamefont {E.}~\bibnamefont
  {Montangero}},\ and\ \bibinfo {author} {\bibnamefont {Evenson}},\ }\href@noop
  {} {\emph {\bibinfo {title} {Introduction to tensor network methods}}}\
  (\bibinfo  {publisher} {Springer},\ \bibinfo {year} {2018})\BibitemShut
  {NoStop}%
\bibitem [{\citenamefont {Ba{\~n}uls}(2023)}]{banuls2023tensor}%
  \BibitemOpen
  \bibfield  {author} {\bibinfo {author} {\bibfnamefont {M.~C.}\ \bibnamefont
  {Ba{\~n}uls}},\ }\bibfield  {title} {\bibinfo {title} {Tensor network
  algorithms: A route map},\ }\href@noop {} {\bibfield  {journal} {\bibinfo
  {journal} {Annual Review of Condensed Matter Physics}\ }\textbf {\bibinfo
  {volume} {14}},\ \bibinfo {pages} {173} (\bibinfo {year} {2023})}\BibitemShut
  {NoStop}%
\bibitem [{\citenamefont {Ran}\ \emph {et~al.}(2020)\citenamefont {Ran},
  \citenamefont {Tirrito}, \citenamefont {Peng}, \citenamefont {Chen},
  \citenamefont {Tagliacozzo}, \citenamefont {Su},\ and\ \citenamefont
  {Lewenstein}}]{ran2020tensor}%
  \BibitemOpen
  \bibfield  {author} {\bibinfo {author} {\bibfnamefont {S.-J.}\ \bibnamefont
  {Ran}}, \bibinfo {author} {\bibfnamefont {E.}~\bibnamefont {Tirrito}},
  \bibinfo {author} {\bibfnamefont {C.}~\bibnamefont {Peng}}, \bibinfo {author}
  {\bibfnamefont {X.}~\bibnamefont {Chen}}, \bibinfo {author} {\bibfnamefont
  {L.}~\bibnamefont {Tagliacozzo}}, \bibinfo {author} {\bibfnamefont
  {G.}~\bibnamefont {Su}},\ and\ \bibinfo {author} {\bibfnamefont
  {M.}~\bibnamefont {Lewenstein}},\ }\href@noop {} {\emph {\bibinfo {title}
  {Tensor network contractions: methods and applications to quantum many-body
  systems}}}\ (\bibinfo  {publisher} {Springer Nature},\ \bibinfo {year}
  {2020})\BibitemShut {NoStop}%
\bibitem [{\citenamefont {Fishman}\ \emph {et~al.}(2022)\citenamefont
  {Fishman}, \citenamefont {White},\ and\ \citenamefont
  {Stoudenmire}}]{fishman2022itensor}%
  \BibitemOpen
  \bibfield  {author} {\bibinfo {author} {\bibfnamefont {M.}~\bibnamefont
  {Fishman}}, \bibinfo {author} {\bibfnamefont {S.}~\bibnamefont {White}},\
  and\ \bibinfo {author} {\bibfnamefont {E.~M.}\ \bibnamefont {Stoudenmire}},\
  }\bibfield  {title} {\bibinfo {title} {The itensor software library for
  tensor network calculations},\ }\href@noop {} {\bibfield  {journal} {\bibinfo
   {journal} {SciPost Physics Codebases}\ ,\ \bibinfo {pages} {004}} (\bibinfo
  {year} {2022})}\BibitemShut {NoStop}%
\bibitem [{\citenamefont {Pan}\ and\ \citenamefont {Meng}(2022)}]{pan2022sign}%
  \BibitemOpen
  \bibfield  {author} {\bibinfo {author} {\bibfnamefont {G.}~\bibnamefont
  {Pan}}\ and\ \bibinfo {author} {\bibfnamefont {Z.~Y.}\ \bibnamefont {Meng}},\
  }\bibfield  {title} {\bibinfo {title} {Sign problem in quantum monte carlo
  simulation},\ }\href@noop {} {\bibfield  {journal} {\bibinfo  {journal}
  {arXiv preprint arXiv:2204.08777}\ } (\bibinfo {year} {2022})}\BibitemShut
  {NoStop}%
\bibitem [{\citenamefont {Troyer}\ and\ \citenamefont
  {Wiese}(2005)}]{troyer2005computational}%
  \BibitemOpen
  \bibfield  {author} {\bibinfo {author} {\bibfnamefont {M.}~\bibnamefont
  {Troyer}}\ and\ \bibinfo {author} {\bibfnamefont {U.-J.}\ \bibnamefont
  {Wiese}},\ }\bibfield  {title} {\bibinfo {title} {Computational complexity
  and fundamental limitations<? format?> to fermionic quantum monte carlo
  simulations},\ }\href@noop {} {\bibfield  {journal} {\bibinfo  {journal}
  {Physical review letters}\ }\textbf {\bibinfo {volume} {94}},\ \bibinfo
  {pages} {170201} (\bibinfo {year} {2005})}\BibitemShut {NoStop}%
\bibitem [{\citenamefont {Schollw\"ock}(2005)}]{SchollwockDMRG}%
  \BibitemOpen
  \bibfield  {author} {\bibinfo {author} {\bibfnamefont {U.}~\bibnamefont
  {Schollw\"ock}},\ }\bibfield  {title} {\bibinfo {title} {The density-matrix
  renormalization group},\ }\href {https://doi.org/10.1103/RevModPhys.77.259}
  {\bibfield  {journal} {\bibinfo  {journal} {Rev. Mod. Phys.}\ }\textbf
  {\bibinfo {volume} {77}},\ \bibinfo {pages} {259} (\bibinfo {year}
  {2005})}\BibitemShut {NoStop}%
\bibitem [{\citenamefont {Tilly}\ \emph {et~al.}(2022)\citenamefont {Tilly},
  \citenamefont {Chen}, \citenamefont {Cao}, \citenamefont {Picozzi},
  \citenamefont {Setia}, \citenamefont {Li}, \citenamefont {Grant},
  \citenamefont {Wossnig}, \citenamefont {Rungger}, \citenamefont {Booth} \emph
  {et~al.}}]{tilly2022variational}%
  \BibitemOpen
  \bibfield  {author} {\bibinfo {author} {\bibfnamefont {J.}~\bibnamefont
  {Tilly}}, \bibinfo {author} {\bibfnamefont {H.}~\bibnamefont {Chen}},
  \bibinfo {author} {\bibfnamefont {S.}~\bibnamefont {Cao}}, \bibinfo {author}
  {\bibfnamefont {D.}~\bibnamefont {Picozzi}}, \bibinfo {author} {\bibfnamefont
  {K.}~\bibnamefont {Setia}}, \bibinfo {author} {\bibfnamefont
  {Y.}~\bibnamefont {Li}}, \bibinfo {author} {\bibfnamefont {E.}~\bibnamefont
  {Grant}}, \bibinfo {author} {\bibfnamefont {L.}~\bibnamefont {Wossnig}},
  \bibinfo {author} {\bibfnamefont {I.}~\bibnamefont {Rungger}}, \bibinfo
  {author} {\bibfnamefont {G.~H.}\ \bibnamefont {Booth}}, \emph {et~al.},\
  }\bibfield  {title} {\bibinfo {title} {The variational quantum eigensolver: a
  review of methods and best practices},\ }\href@noop {} {\bibfield  {journal}
  {\bibinfo  {journal} {Physics Reports}\ }\textbf {\bibinfo {volume} {986}},\
  \bibinfo {pages} {1} (\bibinfo {year} {2022})}\BibitemShut {NoStop}%
\bibitem [{\citenamefont {Kandala}\ \emph {et~al.}(2017)\citenamefont
  {Kandala}, \citenamefont {Mezzacapo}, \citenamefont {Temme}, \citenamefont
  {Takita}, \citenamefont {Brink}, \citenamefont {Chow},\ and\ \citenamefont
  {Gambetta}}]{kandala2017hardware}%
  \BibitemOpen
  \bibfield  {author} {\bibinfo {author} {\bibfnamefont {A.}~\bibnamefont
  {Kandala}}, \bibinfo {author} {\bibfnamefont {A.}~\bibnamefont {Mezzacapo}},
  \bibinfo {author} {\bibfnamefont {K.}~\bibnamefont {Temme}}, \bibinfo
  {author} {\bibfnamefont {M.}~\bibnamefont {Takita}}, \bibinfo {author}
  {\bibfnamefont {M.}~\bibnamefont {Brink}}, \bibinfo {author} {\bibfnamefont
  {J.~M.}\ \bibnamefont {Chow}},\ and\ \bibinfo {author} {\bibfnamefont
  {J.~M.}\ \bibnamefont {Gambetta}},\ }\bibfield  {title} {\bibinfo {title}
  {Hardware-efficient variational quantum eigensolver for small molecules and
  quantum magnets},\ }\href@noop {} {\bibfield  {journal} {\bibinfo  {journal}
  {nature}\ }\textbf {\bibinfo {volume} {549}},\ \bibinfo {pages} {242}
  (\bibinfo {year} {2017})}\BibitemShut {NoStop}%
\bibitem [{\citenamefont {Liu}\ \emph {et~al.}(2019)\citenamefont {Liu},
  \citenamefont {Zhang}, \citenamefont {Wan},\ and\ \citenamefont
  {Wang}}]{liu2019variational}%
  \BibitemOpen
  \bibfield  {author} {\bibinfo {author} {\bibfnamefont {J.-G.}\ \bibnamefont
  {Liu}}, \bibinfo {author} {\bibfnamefont {Y.-H.}\ \bibnamefont {Zhang}},
  \bibinfo {author} {\bibfnamefont {Y.}~\bibnamefont {Wan}},\ and\ \bibinfo
  {author} {\bibfnamefont {L.}~\bibnamefont {Wang}},\ }\bibfield  {title}
  {\bibinfo {title} {Variational quantum eigensolver with fewer qubits},\
  }\href@noop {} {\bibfield  {journal} {\bibinfo  {journal} {Physical Review
  Research}\ }\textbf {\bibinfo {volume} {1}},\ \bibinfo {pages} {023025}
  (\bibinfo {year} {2019})}\BibitemShut {NoStop}%
\bibitem [{\citenamefont {Zhang}\ \emph {et~al.}(2022)\citenamefont {Zhang},
  \citenamefont {Cincio}, \citenamefont {Negre}, \citenamefont {Czarnik},
  \citenamefont {Coles}, \citenamefont {Anisimov}, \citenamefont {Mniszewski},
  \citenamefont {Tretiak},\ and\ \citenamefont {Dub}}]{zhang2022variational}%
  \BibitemOpen
  \bibfield  {author} {\bibinfo {author} {\bibfnamefont {Y.}~\bibnamefont
  {Zhang}}, \bibinfo {author} {\bibfnamefont {L.}~\bibnamefont {Cincio}},
  \bibinfo {author} {\bibfnamefont {C.~F.}\ \bibnamefont {Negre}}, \bibinfo
  {author} {\bibfnamefont {P.}~\bibnamefont {Czarnik}}, \bibinfo {author}
  {\bibfnamefont {P.~J.}\ \bibnamefont {Coles}}, \bibinfo {author}
  {\bibfnamefont {P.~M.}\ \bibnamefont {Anisimov}}, \bibinfo {author}
  {\bibfnamefont {S.~M.}\ \bibnamefont {Mniszewski}}, \bibinfo {author}
  {\bibfnamefont {S.}~\bibnamefont {Tretiak}},\ and\ \bibinfo {author}
  {\bibfnamefont {P.~A.}\ \bibnamefont {Dub}},\ }\bibfield  {title} {\bibinfo
  {title} {Variational quantum eigensolver with reduced circuit complexity},\
  }\href@noop {} {\bibfield  {journal} {\bibinfo  {journal} {npj Quantum
  Information}\ }\textbf {\bibinfo {volume} {8}},\ \bibinfo {pages} {96}
  (\bibinfo {year} {2022})}\BibitemShut {NoStop}%
\bibitem [{\citenamefont {Peruzzo}\ \emph {et~al.}(2014)\citenamefont
  {Peruzzo}, \citenamefont {McClean}, \citenamefont {Shadbolt}, \citenamefont
  {Yung}, \citenamefont {Zhou}, \citenamefont {Love}, \citenamefont
  {Aspuru-Guzik},\ and\ \citenamefont {O’brien}}]{peruzzo2014variational}%
  \BibitemOpen
  \bibfield  {author} {\bibinfo {author} {\bibfnamefont {A.}~\bibnamefont
  {Peruzzo}}, \bibinfo {author} {\bibfnamefont {J.}~\bibnamefont {McClean}},
  \bibinfo {author} {\bibfnamefont {P.}~\bibnamefont {Shadbolt}}, \bibinfo
  {author} {\bibfnamefont {M.-H.}\ \bibnamefont {Yung}}, \bibinfo {author}
  {\bibfnamefont {X.-Q.}\ \bibnamefont {Zhou}}, \bibinfo {author}
  {\bibfnamefont {P.~J.}\ \bibnamefont {Love}}, \bibinfo {author}
  {\bibfnamefont {A.}~\bibnamefont {Aspuru-Guzik}},\ and\ \bibinfo {author}
  {\bibfnamefont {J.~L.}\ \bibnamefont {O’brien}},\ }\bibfield  {title}
  {\bibinfo {title} {A variational eigenvalue solver on a photonic quantum
  processor},\ }\href@noop {} {\bibfield  {journal} {\bibinfo  {journal}
  {Nature communications}\ }\textbf {\bibinfo {volume} {5}},\ \bibinfo {pages}
  {4213} (\bibinfo {year} {2014})}\BibitemShut {NoStop}%
\bibitem [{\citenamefont {McClean}\ \emph {et~al.}(2018)\citenamefont
  {McClean}, \citenamefont {Boixo}, \citenamefont {Smelyanskiy}, \citenamefont
  {Babbush},\ and\ \citenamefont {Neven}}]{mcclean2018barren}%
  \BibitemOpen
  \bibfield  {author} {\bibinfo {author} {\bibfnamefont {J.~R.}\ \bibnamefont
  {McClean}}, \bibinfo {author} {\bibfnamefont {S.}~\bibnamefont {Boixo}},
  \bibinfo {author} {\bibfnamefont {V.~N.}\ \bibnamefont {Smelyanskiy}},
  \bibinfo {author} {\bibfnamefont {R.}~\bibnamefont {Babbush}},\ and\ \bibinfo
  {author} {\bibfnamefont {H.}~\bibnamefont {Neven}},\ }\bibfield  {title}
  {\bibinfo {title} {Barren plateaus in quantum neural network training
  landscapes},\ }\href@noop {} {\bibfield  {journal} {\bibinfo  {journal}
  {Nature communications}\ }\textbf {\bibinfo {volume} {9}},\ \bibinfo {pages}
  {4812} (\bibinfo {year} {2018})}\BibitemShut {NoStop}%
\bibitem [{\citenamefont {Kitaev}(1995)}]{KitaevQPE}%
  \BibitemOpen
  \bibfield  {author} {\bibinfo {author} {\bibfnamefont {A.}~\bibnamefont
  {Kitaev}},\ }\bibfield  {title} {\bibinfo {title} {Quantum measurements and
  the abelian stabilizer problem},\ }\href@noop {} {\bibfield  {journal}
  {\bibinfo  {journal} {arXiv:quant-ph/9511026}\ } (\bibinfo {year}
  {1995})}\BibitemShut {NoStop}%
\bibitem [{\citenamefont {Nielsen}\ and\ \citenamefont
  {Chuang}(2000)}]{nielsen2000qci}%
  \BibitemOpen
  \bibfield  {author} {\bibinfo {author} {\bibfnamefont {M.~A.}\ \bibnamefont
  {Nielsen}}\ and\ \bibinfo {author} {\bibfnamefont {I.~L.}\ \bibnamefont
  {Chuang}},\ }\href@noop {} {\emph {\bibinfo {title} {Quantum Computation and
  Quantum Information}}}\ (\bibinfo  {publisher} {Cambridge University Press},\
  \bibinfo {year} {2000})\BibitemShut {NoStop}%
\bibitem [{\citenamefont {Yoshioka}\ \emph {et~al.}(2025)\citenamefont
  {Yoshioka}, \citenamefont {Amico}, \citenamefont {Kirby}, \citenamefont
  {Jurcevic}, \citenamefont {Dutt}, \citenamefont {Fuller}, \citenamefont
  {Garion}, \citenamefont {Haas}, \citenamefont {Hamamura}, \citenamefont
  {Ivrii} \emph {et~al.}}]{yoshioka2025krylov}%
  \BibitemOpen
  \bibfield  {author} {\bibinfo {author} {\bibfnamefont {N.}~\bibnamefont
  {Yoshioka}}, \bibinfo {author} {\bibfnamefont {M.}~\bibnamefont {Amico}},
  \bibinfo {author} {\bibfnamefont {W.}~\bibnamefont {Kirby}}, \bibinfo
  {author} {\bibfnamefont {P.}~\bibnamefont {Jurcevic}}, \bibinfo {author}
  {\bibfnamefont {A.}~\bibnamefont {Dutt}}, \bibinfo {author} {\bibfnamefont
  {B.}~\bibnamefont {Fuller}}, \bibinfo {author} {\bibfnamefont
  {S.}~\bibnamefont {Garion}}, \bibinfo {author} {\bibfnamefont
  {H.}~\bibnamefont {Haas}}, \bibinfo {author} {\bibfnamefont {I.}~\bibnamefont
  {Hamamura}}, \bibinfo {author} {\bibfnamefont {A.}~\bibnamefont {Ivrii}},
  \emph {et~al.},\ }\bibfield  {title} {\bibinfo {title} {Krylov
  diagonalization of large many-body hamiltonians on a quantum processor},\
  }\href@noop {} {\bibfield  {journal} {\bibinfo  {journal} {Nature
  Communications}\ }\textbf {\bibinfo {volume} {16}},\ \bibinfo {pages} {5014}
  (\bibinfo {year} {2025})}\BibitemShut {NoStop}%
\bibitem [{\citenamefont {Yu}\ \emph {et~al.}(2025{\natexlab{a}})\citenamefont
  {Yu}, \citenamefont {Moreno}, \citenamefont {Iosue}, \citenamefont {Bertels},
  \citenamefont {Claudino}, \citenamefont {Fuller}, \citenamefont
  {Groszkowski}, \citenamefont {Humble}, \citenamefont {Jurcevic},
  \citenamefont {Kirby} \emph {et~al.}}]{yu2025quantum}%
  \BibitemOpen
  \bibfield  {author} {\bibinfo {author} {\bibfnamefont {J.}~\bibnamefont
  {Yu}}, \bibinfo {author} {\bibfnamefont {J.~R.}\ \bibnamefont {Moreno}},
  \bibinfo {author} {\bibfnamefont {J.~T.}\ \bibnamefont {Iosue}}, \bibinfo
  {author} {\bibfnamefont {L.}~\bibnamefont {Bertels}}, \bibinfo {author}
  {\bibfnamefont {D.}~\bibnamefont {Claudino}}, \bibinfo {author}
  {\bibfnamefont {B.}~\bibnamefont {Fuller}}, \bibinfo {author} {\bibfnamefont
  {P.}~\bibnamefont {Groszkowski}}, \bibinfo {author} {\bibfnamefont {T.~S.}\
  \bibnamefont {Humble}}, \bibinfo {author} {\bibfnamefont {P.}~\bibnamefont
  {Jurcevic}}, \bibinfo {author} {\bibfnamefont {W.}~\bibnamefont {Kirby}},
  \emph {et~al.},\ }\bibfield  {title} {\bibinfo {title} {Quantum-centric
  algorithm for sample-based krylov diagonalization},\ }\href@noop {}
  {\bibfield  {journal} {\bibinfo  {journal} {arXiv preprint arXiv:2501.09702}\
  } (\bibinfo {year} {2025}{\natexlab{a}})}\BibitemShut {NoStop}%
\bibitem [{\citenamefont {Brooks}\ \emph {et~al.}(2026)\citenamefont {Brooks},
  \citenamefont {Zou},\ and\ \citenamefont {Rhone}}]{brooks2026ground}%
  \BibitemOpen
  \bibfield  {author} {\bibinfo {author} {\bibfnamefont {C.}~\bibnamefont
  {Brooks}}, \bibinfo {author} {\bibfnamefont {H.}~\bibnamefont {Zou}},\ and\
  \bibinfo {author} {\bibfnamefont {T.~D.}\ \bibnamefont {Rhone}},\ }\bibfield
  {title} {\bibinfo {title} {Ground-state estimation of the heisenberg model on
  frustrated lattices with sample-based krylov quantum diagonalization},\
  }\href@noop {} {\bibfield  {journal} {\bibinfo  {journal} {arXiv preprint
  arXiv:2605.29521}\ } (\bibinfo {year} {2026})}\BibitemShut {NoStop}%
\bibitem [{\citenamefont {et~al.}(2020)}]{MottaITE}%
  \BibitemOpen
  \bibfield  {author} {\bibinfo {author} {\bibfnamefont {M.~M.}\ \bibnamefont
  {et~al.}},\ }\bibfield  {title} {\bibinfo {title} {Determining eigenstates
  and thermal states on a quantum computer using quantum imaginary time
  evolution},\ }\href@noop {} {\bibfield  {journal} {\bibinfo  {journal} {Nat.
  Phys.}\ }\textbf {\bibinfo {volume} {16}},\ \bibinfo {pages} {205} (\bibinfo
  {year} {2020})}\BibitemShut {NoStop}%
\bibitem [{\citenamefont {et~al.}(2019)}]{McArdleITE}%
  \BibitemOpen
  \bibfield  {author} {\bibinfo {author} {\bibfnamefont {S.~M.}\ \bibnamefont
  {et~al.}},\ }\bibfield  {title} {\bibinfo {title} {Variational ansatz-based
  quantum simulation of imaginary time evolution},\ }\href@noop {} {\bibfield
  {journal} {\bibinfo  {journal} {npj Quantum Inf.}\ }\textbf {\bibinfo
  {volume} {5}},\ \bibinfo {pages} {75} (\bibinfo {year} {2019})}\BibitemShut
  {NoStop}%
\bibitem [{\citenamefont {et~al.}(2018)}]{ChildsTrotter}%
  \BibitemOpen
  \bibfield  {author} {\bibinfo {author} {\bibfnamefont {A.~M.~C.}\
  \bibnamefont {et~al.}},\ }\bibfield  {title} {\bibinfo {title} {Toward the
  first quantum simulation with quantum speedup},\ }\href@noop {} {\bibfield
  {journal} {\bibinfo  {journal} {PNAS}\ }\textbf {\bibinfo {volume} {115}},\
  \bibinfo {pages} {9456} (\bibinfo {year} {2018})}\BibitemShut {NoStop}%
\bibitem [{\citenamefont {Campbell}(2019)}]{CampbellTrotter}%
  \BibitemOpen
  \bibfield  {author} {\bibinfo {author} {\bibfnamefont {E.}~\bibnamefont
  {Campbell}},\ }\bibfield  {title} {\bibinfo {title} {Random compiler for fast
  hamiltonian simulation},\ }\href@noop {} {\bibfield  {journal} {\bibinfo
  {journal} {Phys. Rev. Lett.}\ }\textbf {\bibinfo {volume} {123}},\ \bibinfo
  {pages} {070503} (\bibinfo {year} {2019})}\BibitemShut {NoStop}%
\bibitem [{\citenamefont {Kanno}\ \emph {et~al.}(2023)\citenamefont {Kanno},
  \citenamefont {Kohda}, \citenamefont {Imai}, \citenamefont {Koh},
  \citenamefont {Mitarai}, \citenamefont {Mizukami},\ and\ \citenamefont
  {Nakagawa}}]{kanno2023quantum}%
  \BibitemOpen
  \bibfield  {author} {\bibinfo {author} {\bibfnamefont {K.}~\bibnamefont
  {Kanno}}, \bibinfo {author} {\bibfnamefont {M.}~\bibnamefont {Kohda}},
  \bibinfo {author} {\bibfnamefont {R.}~\bibnamefont {Imai}}, \bibinfo {author}
  {\bibfnamefont {S.}~\bibnamefont {Koh}}, \bibinfo {author} {\bibfnamefont
  {K.}~\bibnamefont {Mitarai}}, \bibinfo {author} {\bibfnamefont
  {W.}~\bibnamefont {Mizukami}},\ and\ \bibinfo {author} {\bibfnamefont
  {Y.~O.}\ \bibnamefont {Nakagawa}},\ }\bibfield  {title} {\bibinfo {title}
  {Quantum-selected configuration interaction: Classical diagonalization of
  hamiltonians in subspaces selected by quantum computers},\ }\href@noop {}
  {\bibfield  {journal} {\bibinfo  {journal} {arXiv preprint arXiv:2302.11320}\
  } (\bibinfo {year} {2023})}\BibitemShut {NoStop}%
\bibitem [{\citenamefont {Mikkelsen}\ and\ \citenamefont
  {Nakagawa}(2025)}]{mikkelsen2025quantum}%
  \BibitemOpen
  \bibfield  {author} {\bibinfo {author} {\bibfnamefont {M.}~\bibnamefont
  {Mikkelsen}}\ and\ \bibinfo {author} {\bibfnamefont {Y.~O.}\ \bibnamefont
  {Nakagawa}},\ }\bibfield  {title} {\bibinfo {title} {Quantum-selected
  configuration interaction with time-evolved state},\ }\href@noop {}
  {\bibfield  {journal} {\bibinfo  {journal} {Physical Review Research}\
  }\textbf {\bibinfo {volume} {7}},\ \bibinfo {pages} {043043} (\bibinfo {year}
  {2025})}\BibitemShut {NoStop}%
\bibitem [{\citenamefont {Sugisaki}\ \emph {et~al.}(2025)\citenamefont
  {Sugisaki}, \citenamefont {Kanno}, \citenamefont {Itoko}, \citenamefont
  {Sakuma},\ and\ \citenamefont {Yamamoto}}]{sugisaki2025hamiltonian}%
  \BibitemOpen
  \bibfield  {author} {\bibinfo {author} {\bibfnamefont {K.}~\bibnamefont
  {Sugisaki}}, \bibinfo {author} {\bibfnamefont {S.}~\bibnamefont {Kanno}},
  \bibinfo {author} {\bibfnamefont {T.}~\bibnamefont {Itoko}}, \bibinfo
  {author} {\bibfnamefont {R.}~\bibnamefont {Sakuma}},\ and\ \bibinfo {author}
  {\bibfnamefont {N.}~\bibnamefont {Yamamoto}},\ }\bibfield  {title} {\bibinfo
  {title} {Hamiltonian simulation-based quantum-selected configuration
  interaction for large-scale electronic structure calculations with a quantum
  computer},\ }\href@noop {} {\bibfield  {journal} {\bibinfo  {journal}
  {Physical Chemistry Chemical Physics}\ }\textbf {\bibinfo {volume} {27}},\
  \bibinfo {pages} {20869} (\bibinfo {year} {2025})}\BibitemShut {NoStop}%
\bibitem [{\citenamefont {Nakagawa}\ \emph {et~al.}(2024)\citenamefont
  {Nakagawa}, \citenamefont {Kamoshita}, \citenamefont {Mizukami},
  \citenamefont {Sudo},\ and\ \citenamefont {Ohnishi}}]{nakagawa2024adapt}%
  \BibitemOpen
  \bibfield  {author} {\bibinfo {author} {\bibfnamefont {Y.~O.}\ \bibnamefont
  {Nakagawa}}, \bibinfo {author} {\bibfnamefont {M.}~\bibnamefont {Kamoshita}},
  \bibinfo {author} {\bibfnamefont {W.}~\bibnamefont {Mizukami}}, \bibinfo
  {author} {\bibfnamefont {S.}~\bibnamefont {Sudo}},\ and\ \bibinfo {author}
  {\bibfnamefont {Y.-y.}\ \bibnamefont {Ohnishi}},\ }\bibfield  {title}
  {\bibinfo {title} {Adapt-qsci: Adaptive construction of an input state for
  quantum-selected configuration interaction},\ }\href@noop {} {\bibfield
  {journal} {\bibinfo  {journal} {Journal of Chemical Theory and Computation}\
  }\textbf {\bibinfo {volume} {20}},\ \bibinfo {pages} {10817} (\bibinfo {year}
  {2024})}\BibitemShut {NoStop}%
\bibitem [{\citenamefont {Patra}\ \emph {et~al.}(2025)\citenamefont {Patra},
  \citenamefont {Mondal}, \citenamefont {Halder}, \citenamefont {Halder},
  \citenamefont {Laskar}, \citenamefont {Goel},\ and\ \citenamefont
  {Maitra}}]{patra2025physics}%
  \BibitemOpen
  \bibfield  {author} {\bibinfo {author} {\bibfnamefont {C.}~\bibnamefont
  {Patra}}, \bibinfo {author} {\bibfnamefont {D.}~\bibnamefont {Mondal}},
  \bibinfo {author} {\bibfnamefont {S.}~\bibnamefont {Halder}}, \bibinfo
  {author} {\bibfnamefont {D.}~\bibnamefont {Halder}}, \bibinfo {author}
  {\bibfnamefont {M.~R.}\ \bibnamefont {Laskar}}, \bibinfo {author}
  {\bibfnamefont {R.}~\bibnamefont {Goel}},\ and\ \bibinfo {author}
  {\bibfnamefont {R.}~\bibnamefont {Maitra}},\ }\bibfield  {title} {\bibinfo
  {title} {Physics-informed generative machine learning for accelerated
  quantum-centric supercomputing},\ }\href@noop {} {\bibfield  {journal}
  {\bibinfo  {journal} {arXiv preprint arXiv:2512.06858}\ } (\bibinfo {year}
  {2025})}\BibitemShut {NoStop}%
\bibitem [{\citenamefont {Giamarchi}(2003)}]{GiamarchiBook}%
  \BibitemOpen
  \bibfield  {author} {\bibinfo {author} {\bibfnamefont {T.}~\bibnamefont
  {Giamarchi}},\ }\href@noop {} {\emph {\bibinfo {title} {Quantum Physics in
  One Dimension}}}\ (\bibinfo  {publisher} {Oxford University Press},\ \bibinfo
  {year} {2003})\BibitemShut {NoStop}%
\bibitem [{\citenamefont {Takahashi}(1999)}]{TakahashiBook}%
  \BibitemOpen
  \bibfield  {author} {\bibinfo {author} {\bibfnamefont {M.}~\bibnamefont
  {Takahashi}},\ }\href@noop {} {\emph {\bibinfo {title} {Thermodynamics of
  One-Dimensional Solvable Models}}}\ (\bibinfo  {publisher} {Cambridge
  University Press},\ \bibinfo {year} {1999})\BibitemShut {NoStop}%
\bibitem [{\citenamefont {Sandvik}(2010{\natexlab{b}})}]{sandvik}%
  \BibitemOpen
  \bibfield  {author} {\bibinfo {author} {\bibfnamefont {A.~W.}\ \bibnamefont
  {Sandvik}},\ }\bibfield  {title} {\bibinfo {title} {Computational studies of
  quantum spin systems},\ }\href {https://doi.org/10.1063/1.3518900} {\bibfield
   {journal} {\bibinfo  {journal} {AIP Conference Proceedings}\ }\textbf
  {\bibinfo {volume} {1297}},\ \bibinfo {pages} {135} (\bibinfo {year}
  {2010}{\natexlab{b}})}\BibitemShut {NoStop}%
\bibitem [{\citenamefont {Rahaman}\ \emph {et~al.}(2023)\citenamefont
  {Rahaman}, \citenamefont {Haldar},\ and\ \citenamefont
  {Kumar}}]{rahaman2023machine}%
  \BibitemOpen
  \bibfield  {author} {\bibinfo {author} {\bibfnamefont {S.~S.}\ \bibnamefont
  {Rahaman}}, \bibinfo {author} {\bibfnamefont {S.}~\bibnamefont {Haldar}},\
  and\ \bibinfo {author} {\bibfnamefont {M.}~\bibnamefont {Kumar}},\ }\bibfield
   {title} {\bibinfo {title} {Machine learning approach to study quantum phase
  transitions of a frustrated one dimensional spin-1/2 system},\ }\href@noop {}
  {\bibfield  {journal} {\bibinfo  {journal} {Journal of Physics: Condensed
  Matter}\ }\textbf {\bibinfo {volume} {35}},\ \bibinfo {pages} {115603}
  (\bibinfo {year} {2023})}\BibitemShut {NoStop}%
\bibitem [{\citenamefont {Robledo-Moreno}\ \emph {et~al.}(2025)\citenamefont
  {Robledo-Moreno}, \citenamefont {Motta}, \citenamefont {Haas}, \citenamefont
  {Javadi-Abhari}, \citenamefont {Jurcevic}, \citenamefont {Kirby},
  \citenamefont {Martiel}, \citenamefont {Sharma}, \citenamefont {Sharma},
  \citenamefont {Shirakawa}, \citenamefont {Sitdikov}, \citenamefont {Sun},
  \citenamefont {Sung}, \citenamefont {Takita}, \citenamefont {Tran},
  \citenamefont {Yunoki},\ and\ \citenamefont {Mezzacapo}}]{RobMottaSQD}%
  \BibitemOpen
  \bibfield  {author} {\bibinfo {author} {\bibfnamefont {J.}~\bibnamefont
  {Robledo-Moreno}}, \bibinfo {author} {\bibfnamefont {M.}~\bibnamefont
  {Motta}}, \bibinfo {author} {\bibfnamefont {H.}~\bibnamefont {Haas}},
  \bibinfo {author} {\bibfnamefont {A.}~\bibnamefont {Javadi-Abhari}}, \bibinfo
  {author} {\bibfnamefont {P.}~\bibnamefont {Jurcevic}}, \bibinfo {author}
  {\bibfnamefont {W.}~\bibnamefont {Kirby}}, \bibinfo {author} {\bibfnamefont
  {S.}~\bibnamefont {Martiel}}, \bibinfo {author} {\bibfnamefont
  {K.}~\bibnamefont {Sharma}}, \bibinfo {author} {\bibfnamefont
  {S.}~\bibnamefont {Sharma}}, \bibinfo {author} {\bibfnamefont
  {T.}~\bibnamefont {Shirakawa}}, \bibinfo {author} {\bibfnamefont
  {I.}~\bibnamefont {Sitdikov}}, \bibinfo {author} {\bibfnamefont {R.-Y.}\
  \bibnamefont {Sun}}, \bibinfo {author} {\bibfnamefont {K.~J.}\ \bibnamefont
  {Sung}}, \bibinfo {author} {\bibfnamefont {M.}~\bibnamefont {Takita}},
  \bibinfo {author} {\bibfnamefont {M.~C.}\ \bibnamefont {Tran}}, \bibinfo
  {author} {\bibfnamefont {S.}~\bibnamefont {Yunoki}},\ and\ \bibinfo {author}
  {\bibfnamefont {A.}~\bibnamefont {Mezzacapo}},\ }\bibfield  {title} {\bibinfo
  {title} {Chemistry beyond the scale of exact diagonalization on a
  quantum-centric supercomputer},\ }\href
  {https://doi.org/10.1126/sciadv.adu9991} {\bibfield  {journal} {\bibinfo
  {journal} {Science Advances}\ }\textbf {\bibinfo {volume} {11}},\ \bibinfo
  {pages} {eadu9991} (\bibinfo {year} {2025})}\BibitemShut {NoStop}%
\bibitem [{\citenamefont {Misciasci}\ \emph {et~al.}(2026)\citenamefont
  {Misciasci}, \citenamefont {Firt}, \citenamefont {Mueller}, \citenamefont
  {Friedhoff}, \citenamefont {Onah}, \citenamefont {Schulze},\ and\
  \citenamefont {Mostame}}]{misciasci2026}%
  \BibitemOpen
  \bibfield  {author} {\bibinfo {author} {\bibfnamefont {N.}~\bibnamefont
  {Misciasci}}, \bibinfo {author} {\bibfnamefont {R.}~\bibnamefont {Firt}},
  \bibinfo {author} {\bibfnamefont {J.~E.}\ \bibnamefont {Mueller}}, \bibinfo
  {author} {\bibfnamefont {T.}~\bibnamefont {Friedhoff}}, \bibinfo {author}
  {\bibfnamefont {C.}~\bibnamefont {Onah}}, \bibinfo {author} {\bibfnamefont
  {A.}~\bibnamefont {Schulze}},\ and\ \bibinfo {author} {\bibfnamefont
  {S.}~\bibnamefont {Mostame}},\ }\bibfield  {title} {\bibinfo {title}
  {Evaluating sample-based krylov quantum diagonalization for heisenberg models
  with applications to materials science},\ }\bibfield  {journal} {\bibinfo
  {journal} {Entropy}\ }\textbf {\bibinfo {volume} {28}},\ \href
  {https://doi.org/10.3390/e28040367} {10.3390/e28040367} (\bibinfo {year}
  {2026})\BibitemShut {NoStop}%
\bibitem [{\citenamefont {Yu}\ \emph {et~al.}(2025{\natexlab{b}})\citenamefont
  {Yu}, \citenamefont {Robledo~Moreno}, \citenamefont {Iosue}, \citenamefont
  {Amico}, \citenamefont {Bertels}, \citenamefont {Claudino}, \citenamefont
  {Fuller}, \citenamefont {Groszkowski}, \citenamefont {Humble}, \citenamefont
  {Jurcevic}, \citenamefont {Kirby}, \citenamefont {Maier}, \citenamefont
  {Motta}, \citenamefont {Pokharel}, \citenamefont {Seif}, \citenamefont
  {Shehata}, \citenamefont {Sung}, \citenamefont {Tran}, \citenamefont
  {Tripathi}, \citenamefont {Mezzacapo},\ and\ \citenamefont
  {Sharma}}]{YuSKQD}%
  \BibitemOpen
  \bibfield  {author} {\bibinfo {author} {\bibfnamefont {J.}~\bibnamefont
  {Yu}}, \bibinfo {author} {\bibfnamefont {J.}~\bibnamefont {Robledo~Moreno}},
  \bibinfo {author} {\bibfnamefont {J.~T.}\ \bibnamefont {Iosue}}, \bibinfo
  {author} {\bibfnamefont {M.}~\bibnamefont {Amico}}, \bibinfo {author}
  {\bibfnamefont {L.}~\bibnamefont {Bertels}}, \bibinfo {author} {\bibfnamefont
  {D.}~\bibnamefont {Claudino}}, \bibinfo {author} {\bibfnamefont
  {B.}~\bibnamefont {Fuller}}, \bibinfo {author} {\bibfnamefont
  {P.}~\bibnamefont {Groszkowski}}, \bibinfo {author} {\bibfnamefont {T.~S.}\
  \bibnamefont {Humble}}, \bibinfo {author} {\bibfnamefont {P.}~\bibnamefont
  {Jurcevic}}, \bibinfo {author} {\bibfnamefont {W.}~\bibnamefont {Kirby}},
  \bibinfo {author} {\bibfnamefont {T.~A.}\ \bibnamefont {Maier}}, \bibinfo
  {author} {\bibfnamefont {M.}~\bibnamefont {Motta}}, \bibinfo {author}
  {\bibfnamefont {B.}~\bibnamefont {Pokharel}}, \bibinfo {author}
  {\bibfnamefont {A.}~\bibnamefont {Seif}}, \bibinfo {author} {\bibfnamefont
  {A.}~\bibnamefont {Shehata}}, \bibinfo {author} {\bibfnamefont {K.~J.}\
  \bibnamefont {Sung}}, \bibinfo {author} {\bibfnamefont {M.~C.}\ \bibnamefont
  {Tran}}, \bibinfo {author} {\bibfnamefont {V.}~\bibnamefont {Tripathi}},
  \bibinfo {author} {\bibfnamefont {A.}~\bibnamefont {Mezzacapo}},\ and\
  \bibinfo {author} {\bibfnamefont {K.}~\bibnamefont {Sharma}},\ }\bibfield
  {title} {\bibinfo {title} {Quantum-centric algorithm for sample-based
  {K}rylov diagonalization},\ }\href@noop {} {\bibfield  {journal} {\bibinfo
  {journal} {arXiv preprint arXiv:2501.09702}\ } (\bibinfo {year}
  {2025}{\natexlab{b}})}\BibitemShut {NoStop}%
\bibitem [{\citenamefont {Epperly}\ \emph {et~al.}(2022)\citenamefont
  {Epperly}, \citenamefont {Lin},\ and\ \citenamefont
  {Nakatsukasa}}]{EpperlyKQD}%
  \BibitemOpen
  \bibfield  {author} {\bibinfo {author} {\bibfnamefont {E.~N.}\ \bibnamefont
  {Epperly}}, \bibinfo {author} {\bibfnamefont {L.}~\bibnamefont {Lin}},\ and\
  \bibinfo {author} {\bibfnamefont {Y.}~\bibnamefont {Nakatsukasa}},\
  }\bibfield  {title} {\bibinfo {title} {A theory of quantum subspace
  diagonalization},\ }\href {https://doi.org/10.1137/21M145954X} {\bibfield
  {journal} {\bibinfo  {journal} {SIAM Journal on Matrix Analysis and
  Applications}\ }\textbf {\bibinfo {volume} {43}},\ \bibinfo {pages} {1263}
  (\bibinfo {year} {2022})}\BibitemShut {NoStop}%
\bibitem [{\citenamefont {Hastings}(2007)}]{HastingsAreaLaw}%
  \BibitemOpen
  \bibfield  {author} {\bibinfo {author} {\bibfnamefont {M.~B.}\ \bibnamefont
  {Hastings}},\ }\bibfield  {title} {\bibinfo {title} {An area law for
  one-dimensional quantum systems},\ }\href
  {https://doi.org/10.1088/1742-5468/2007/08/P08024} {\bibfield  {journal}
  {\bibinfo  {journal} {Journal of Statistical Mechanics: Theory and
  Experiment}\ }\textbf {\bibinfo {volume} {2007}},\ \bibinfo {pages} {P08024}
  (\bibinfo {year} {2007})}\BibitemShut {NoStop}%
\bibitem [{\citenamefont {Eisert}\ \emph {et~al.}(2010)\citenamefont {Eisert},
  \citenamefont {Cramer},\ and\ \citenamefont {Plenio}}]{EisertAreaLaw}%
  \BibitemOpen
  \bibfield  {author} {\bibinfo {author} {\bibfnamefont {J.}~\bibnamefont
  {Eisert}}, \bibinfo {author} {\bibfnamefont {M.}~\bibnamefont {Cramer}},\
  and\ \bibinfo {author} {\bibfnamefont {M.~B.}\ \bibnamefont {Plenio}},\
  }\bibfield  {title} {\bibinfo {title} {Colloquium: Area laws for the
  entanglement entropy},\ }\href {https://doi.org/10.1103/RevModPhys.82.277}
  {\bibfield  {journal} {\bibinfo  {journal} {Reviews of Modern Physics}\
  }\textbf {\bibinfo {volume} {82}},\ \bibinfo {pages} {277} (\bibinfo {year}
  {2010})}\BibitemShut {NoStop}%
\end{thebibliography}%

%============================================================
\appendix
%============================================================
\section{Trotterized Quantum Circuit for the XXZ Hamiltonian}
\label{app:trotter}

\begin{figure}[h!]
    \centering
    \includegraphics[width=0.95\linewidth]{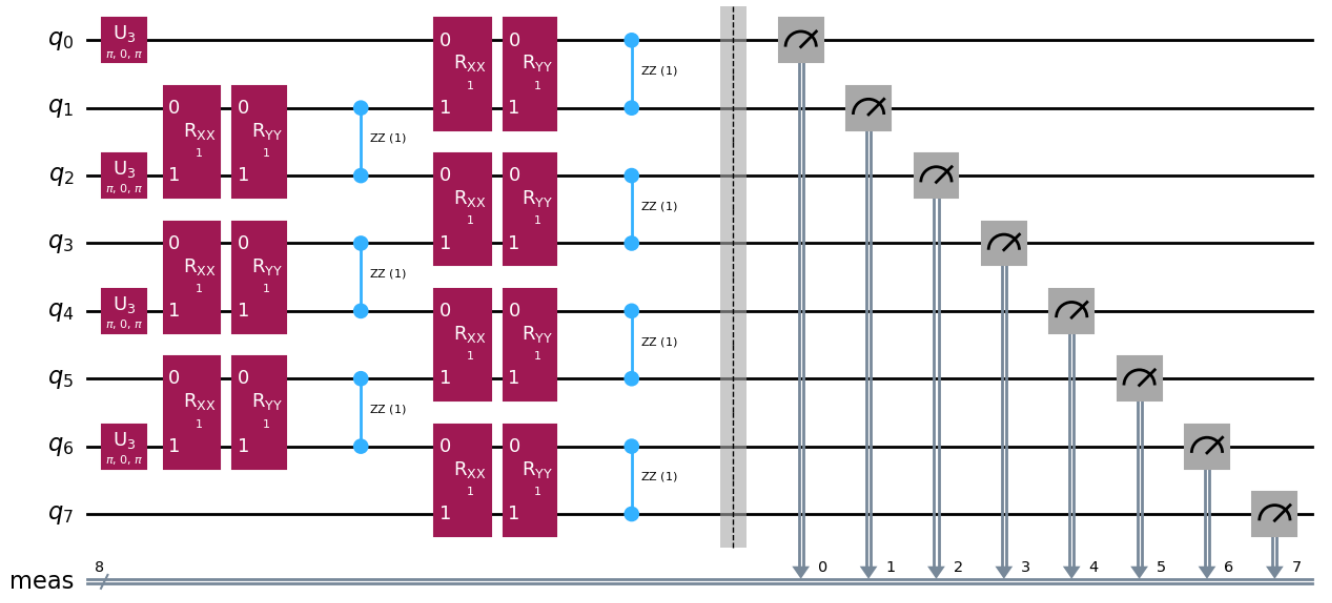}
    \caption{
    Trotterized quantum circuit implementing one Trotter step of the XXZ Hamiltonian (Eq.~(1)) using the first-order Lie--Trotter decomposition. 
    The $R_{XX}$, $R_{YY}$, and $ZZ$ rotation gates implement the $XY$-exchange and Ising interactions on each bond. 
    Initial single-qubit $U_3$ rotations prepare the input computational-basis configuration.
    }
    \label{fig:supp_trotter}
\end{figure}

In the Basis-Adaptive (BA) algorithm described in the main text, the central quantum subroutine is the application of the unitary operator
\[
U(\Delta t) = e^{-iH\Delta t},
\]
to each basis configuration used as an independent sample wavefunction.  
This supplementary note explains the construction of the Trotterized quantum circuit that implements $U(\Delta t)$ for the XXZ Hamiltonian. A representative circuit is shown in Fig.~\ref{fig:supp_trotter}.

\subsection*{A1. Decomposition of the XXZ Hamiltonian}

The nearest-neighbour XXZ Hamiltonian is
\[
H = \sum_j H_j,
\qquad
H_j = J(S_j^x S_{j+1}^x + S_j^y S_{j+1}^y) + \Delta\, S_j^z S_{j+1}^z.
\]

Each local term naturally separates into two non-commuting pieces:
\[
H_j = H_{XY}^{(j)} + H_{ZZ}^{(j)},
\]
where
\[
H_{XY}^{(j)} = J(X_j X_{j+1} + Y_j Y_{j+1}), \qquad 
H_{ZZ}^{(j)} = \Delta Z_j Z_{j+1}.
\]

Because the two parts do not commute,
\(
[H_{XY}^{(j)}, H_{ZZ}^{(j)}] \neq 0
\),
the exact evolution operator is approximated using the first-order Lie--Trotter formula:
\begin{equation}
e^{-iH\Delta t} 
\approx 
\prod_{j} 
e^{-iH_{XY}^{(j)}\Delta t}\,
e^{-iH_{ZZ}^{(j)}\Delta t}
+ 
\mathcal{O}((\Delta t)^2).
\label{eq:trotter_supp}
\end{equation}

This factorization directly determines the layered structure of the quantum circuit.

\subsection*{A2. Implementation Using Entangling Gates}

\subsubsection*{(a) XY exchange: $R_{XX}$ and $R_{YY}$ gates}

For each bond $(j,j+1)$, the XY part evolves under
\[
e^{-i H_{XY}^{(j)} \Delta t}
=
e^{-i J\Delta t\, X_j X_{j+1}}
\, e^{-i J\Delta t\, Y_j Y_{j+1}}.
\]

The corresponding two-qubit rotations are implemented using
\[
R_{XX}(\theta) = e^{-i\frac{\theta}{2} X\otimes X}, 
\qquad
R_{YY}(\theta) = e^{-i\frac{\theta}{2} Y\otimes Y},
\]
with rotation angle
\[
\frac{\theta}{2} = J\Delta t.
\]

Sequential application of $R_{XX}$ and $R_{YY}$ realizes the complete $XY$ exchange for one Trotter slice.

\subsubsection*{(b) Ising term: $ZZ$ rotation}

The longitudinal interaction contributes
\[
e^{-iH_{ZZ}^{(j)}\Delta t}
= 
e^{-iJ \Delta\Delta t\, Z_j Z_{j+1}}
\equiv ZZ(\phi),
\]
with 
\[
\frac{\phi}{2} = J \Delta\,\Delta t.
\]

Most hardware platforms support $ZZ(\theta)$ natively, or via a controlled-phase gate decomposition.

\subsubsection*{(d) Trotter Layer: Gate Decomposition}

Their decompositions into basis gates are:

\paragraph{Decomposition of \(R_{ZZ}(\theta)\):}
\[
R_{ZZ}(\theta) = 
\begin{quantikz}
\lstick{} & \ctrl{1} & \gate{R_z(\theta)} & \ctrl{1} & \rstick{} \\
\lstick{} & \targ{}  & \qw               & \targ{}  & \rstick{}
\end{quantikz}
\]

\paragraph{Decomposition of \(R_{XX}(\theta)\):}
\[
R_{XX}(\theta) = 
(H \otimes H)\,
R_{ZZ}(\theta)\,
(H \otimes H)
\]

\paragraph{Decomposition of \(R_{YY}(\theta)\):}
\[
R_{YY}(\theta) =
(R_x(\tfrac{\pi}{2}) \otimes R_x(\tfrac{\pi}{2}))\,
R_{ZZ}(\theta)\,
(R_x(-\tfrac{\pi}{2}) \otimes R_x(-\tfrac{\pi}{2}))
\]

\subsubsection*{(c) Full Trotter layer}

One complete Trotter layer thus consists of:
\begin{itemize}
    \item $R_{XX}$ rotations on all even bonds,
    \item $R_{YY}$ rotations on all even bonds,
    \item $ZZ$ rotations on all even bonds,
    \item followed by the same three gates on odd bonds.
\end{itemize}

%------------------------------------

\subsection*{A3. Role of the Circuit in the Basis-Adaptive Algorithm}

Each computational-basis configuration $\ket{b_k^{(i)}}$ is treated as an independent sample wavefunction.  
For each such sample, the Trotter circuit implements
\[
\ket{b_k^{(i)'}} = U(\Delta t)\ket{b_k^{(i)}}.
\]

The measurements from all samples are collected to form the union set of bitstrings.  
Crucially:
\begin{itemize}
    \item The full wavefunction is \emph{never reconstructed} on hardware.
    \item Only the sampled computational-basis outputs are retained.
    \item Symmetry filtering enforces conservation of total $S^z$ (U(1) symmetry) and reflection invariance.
\end{itemize}

The filtered set defines the reduced Hilbert subspace for the next iteration, in which we perform an ED calculation to obtain the approximate ground state.
\subsection*{A4. XXZ Hamiltonian in weak perturbation limit }
To understand the underlying mechanism, we consider the Hamiltonian $H$ given in Eq.~1 of the main text. 
For large values of $\Delta$, the model reduces to the pure Ising limit, and the transverse exchange terms 
can be treated as a small perturbation. The Hamiltonian can therefore be written as
\[
H = \sum_i S_i^z S_{i+1}^z
+ \gamma \sum_i \frac{1}{2}\left( S_i^+ S_{i+1}^- + S_i^- S_{i+1}^+ \right),
\]
where $\gamma = 1/\Delta$. In the Ising limit ($\gamma \to 0$), the ground state $|\alpha\rangle$ and one of the 
two--spin--flip excited states $|\alpha'\rangle$ are
\[
|\alpha\rangle = |\uparrow \downarrow \uparrow \downarrow \cdots\rangle,
\qquad
|\alpha'\rangle = |\uparrow \downarrow \uparrow \uparrow \downarrow \downarrow \cdots\rangle.
\]

The energy of the Ising ground state is
\[
E_0 = -\frac{N}{4},
\]
while the two--spin--flip excitation has energy
\[
E_0^{\alpha'} = -\frac{N}{4} + 1 
= -\frac{1}{4}(N - 4).
\]

The second--order correction to the ground--state energy arises from these two--spin--flip states:
\[
\Delta E^{(2)}
= -\sum_{\alpha'}
\frac{|\langle \alpha' | H' | \alpha \rangle|^2}
     {E^{\alpha'} - E^\alpha}.
\]
Since there are $N$ such states, this simplifies to
\[
\Delta E^{(2)} = -N \left(\frac{1}{2}\gamma\right)^2
= -\frac{N}{4}\gamma^2 .
\]

At fourth order, contributions come from states involving four spin flips. 
On a periodic chain, there are $N(N-1)/2$ such configurations, giving
\[
\Delta E^{(4)} \propto \frac{(N-1)N}{2}\,\gamma^4 .
\]

Thus, for small $\gamma$, only the two--spin--flip and four--spin--flip sectors contribute significantly to the 
perturbative energy, while higher--order flip processes are negligible. 
This implies that during time evolution, the irrelevant high--order spin--flip sectors may be discarded at 
each step to retain only the physically relevant degrees of freedom. The percentage error in gs energy is defined as, $\Delta E_{gs}=\frac{|E_{gs}(Perturbative)-E_{gs}(ED)|}{E_{gs}(ED)} \times 100$.

\begin{table}[h!]
\centering
\caption{Comparison of exact ground-state energy and perturbative energy for different $\gamma$ values}
\label{tab:gs_compare}
\begin{tabular}{|c|c|c|c|c|}
\hline
$\Delta$ & Exact $E_{\mathrm{GS}}(ED)$ & Perturbative $E_{\mathrm{GS}}$ &  $\Delta E_{gs}$  \\ 
\hline
0.1 & -6.05985 & -6.06 &   0.0025 \\ 
\hline
0.05 & -6.0149 & -6.015 &  0.0016  \\ 
  
\hline
\end{tabular}
\end{table}

\subsection*{A5. Summary}

The Trotterized circuit in Fig.~\ref{fig:supp_trotter} efficiently encodes the $XY$ and $ZZ$ interactions of the XXZ Hamiltonian using native $R_{XX}$, $R_{YY}$, and $ZZ$ gates.  
This circuit forms the core quantum subroutine of the basis-adaptive hybrid algorithm, allowing us to sample physically relevant configurations while avoiding the exponential complexity of a full wavefunction representation.

%--------------------------------------------------------------

\section{Derivation of the General Energy Bound}
\label{app:bounds}

We derive Eq.~\eqref{eq:master_bound} from first principles.

Let $H$ act on an $D_{\mathcal{H}}=2^n$ dimensional Hilbert space.
Expand the true ground state in the computational basis with
coefficients ordered by decreasing magnitude:
\begin{equation}
  |\phi_0\rangle = \sum_{j=1}^{D_{\mathcal{H}}} c_j |b_j\rangle,
  \quad |c_1|\ge|c_2|\ge\cdots\ge|c_{D_{\mathcal{H}}}|.
\end{equation}
The sparsity parameters $(\alpha_{D_T},\beta_{D_T})$ are defined by
\begin{align}
  \alpha_{D_T} &= \sum_{j=1}^{D_T}|c_j|^2 \ge \alpha_{D_T},\\
  |c_j|^2  &\ge \beta_{D_T} \quad \forall\, j=1,\dots,{D_T} . 
\end{align}
The approximate state formed from the $L$ identified bitstrings is
\begin{equation}
  |\tilde\phi\rangle = \frac{1}{C}\sum_{j=1}^{D_T}c_j|b_j\rangle,
  \quad C=\Bigl(\sum_{j=1}^{D_T}|c_j|^2\Bigr)^{1/2}\ge\sqrt{\alpha_{D_T}}.
\end{equation}

No let us define the difference vector $|\phi'\rangle = |\tilde\phi\rangle - |\phi_0\rangle$.
Writing $|\tilde\phi\rangle = |\phi_0\rangle + |\phi'\rangle$ and expanding:
\begin{align}
  \Delta E &= \langle\tilde\phi|H|\tilde\phi\rangle
             - \langle\phi_0|H|\phi_0\rangle \nonumber\\
           &= \langle\tilde\phi|H|\phi'\rangle
             + \langle\phi'|H|\phi_0\rangle .
\end{align}

We can write the Cauchy--Schwarz equations  for  $|\tilde\phi\rangle$ and $|\phi_0\rangle$  which are both normalized then
\begin{equation}
  \Delta E \le 2\|H\|\,\||\phi'\rangle\| .
\end{equation}

and using the above equations we can expand $|\phi'\rangle$ explicitly in the full basis:
\begin{equation}
  |\phi'\rangle = \sum_{j=1}^{L}c_j\!\left(\frac{1}{C}-1\right)|b_j\rangle
               - \sum_{j=L+1}^{D_{\mathcal{H}}}c_j|b_j\rangle .
\end{equation}
Since the two sums run over orthogonal subspaces:
\begin{align}
  \||\phi'\rangle\|^2
    &= \left(\frac{1}{C}-1\right)^{\!2}C^2 + (1-C^2) \nonumber\\
    &= (1-C)^2 + (1-C^2) = 2 - 2C .
\end{align}
Hence $\||\phi'\rangle\| = \sqrt{2}\,\sqrt{1-C}$.

Now using $C\ge\sqrt{\alpha_{D_T}}$, we can write
\begin{equation}
  \||\phi'\rangle\| \le \sqrt{2}\bigl(1-\sqrt{\alpha_{D_T}}\bigr)^{1/2}.
\end{equation}
Substituting into (B6) and noting $2\sqrt{2}=\sqrt{8}$:
\begin{equation}
  \Delta E \le \sqrt{8}\,\|H\|\bigl(1-\sqrt{\alpha_{D_T}}\bigr)^{1/2}.
\end{equation}
This completes the derivation of Eq.~\eqref{eq:master_bound}.

This proof follows closely Lemma~5
(Appendix~B.1) of Ref.~\cite{YuSKQD} and the analogous result in
Ref.~\cite{RobMottaSQD}, both of which build on the Krylov subspace
framework of Ref.~\cite{EpperlyKQD}.

\subsection*{B1. Monotonicity note}

The function $f(\alpha)=(1-\sqrt{\alpha})^{1/2}$ satisfies
$\mathrm{d}f/\mathrm{d}\alpha = -1/(4\sqrt{\alpha}\sqrt{1-\sqrt{\alpha}})<0$
for $\alpha\in(0,1)$, so any lower bound $\alpha_{D_T}\ge\alpha_{\min}$
immediately yields an upper bound on $\Delta E$. This justifies using
phase-specific estimates for $\alpha_L$ derived below.

%------------------------------------------------------------
\section{Ground-State Sparsity Derivations for the XXZ Chain}
\label{app:qlro}
%------------------------------------------------------------

\subsection*{C1. Gapped phase ($\Delta>1$): exponential decay}

In the N\'eel phase ($\Delta>1$), the system possesses a finite
spectral gap and obeys an area law for entanglement
entropy~\cite{HastingsAreaLaw,EisertAreaLaw}. Let $D_{\mathcal H}$ denote the full Hilbert-space dimension and
$D_T$ the number of retained basis states in the BA truncation. After sorting the computational-basis probabilities in descending
order,
\begin{equation}
|c_0|^2 \ge |c_1|^2 \ge |c_2|^2 \ge \cdots ,
\end{equation}
the retained weight is defined as
\begin{equation}
\alpha_{D_T}
=
\sum_{j=0}^{D_T}|c_j|^2.
\end{equation}

\begin{figure}[h]
\includegraphics[width=1.0\columnwidth]{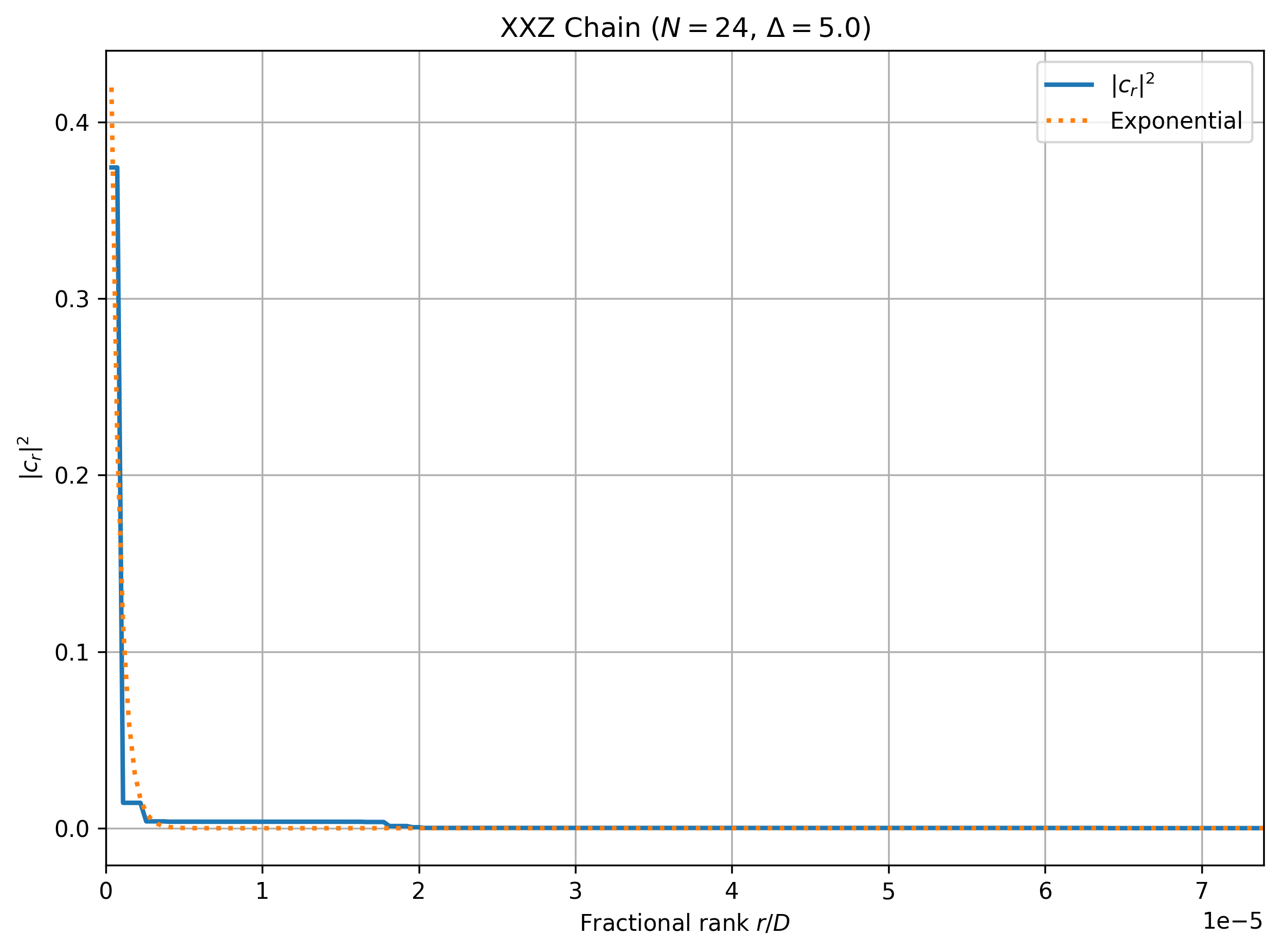}
\caption{For N=24 and $\Delta=5.0$, the computational-basis probabilities $|c_j|^2$ varies exponentially with the Hilbert space $j$.}
\label{fig:corsz_comp}
\end{figure}

To characterize the sparsity of the ground state in the gapped N\'eel
phase, we numerically examined the sorted computational-basis
probabilities $|c_j|^2$. For $N=24$ and $\Delta=5$, the distribution
is well fitted by
\begin{equation}
    |c_j|^2 \simeq c_0 e^{-j/j_0},
\end{equation}
with $j$ denoting the rank of a basis state after sorting in
descending order of weight. We observe qualitatively similar behavior
for other values of $\Delta>1$. Assuming this exponential form, the
retained probability weight $\alpha_{D_T}$ can be estimated
analytically as follows.

The cumulative retained weight of the largest $D_T$ basis states is
\begin{equation}
    \alpha_{D_T}
    =
    \sum_{j=0}^{D_T}|c_j|^2
    \approx
    \int_0^{D_T} c_0 e^{-j/j_0}\,dj.
\end{equation}
Evaluating the integral gives
\begin{equation}
    \alpha_{D_T}
    =
    c_0 j_0
    \left(
      1-e^{-D_T/j_0}
    \right).
\end{equation}

The normalization condition
\begin{equation}
    \sum_{j=0}^{D_{\mathcal H}} |c_j|^2 =1
\end{equation}
yields
\begin{equation}
    1
    =
    \int_0^{D_{\mathcal H}}
    c_0 e^{-j/j_0}\,dj
    =
    c_0 j_0
    \left(
      1-e^{-D_{\mathcal H}/j_0}
    \right),
\end{equation}
so that
\begin{equation}
    c_0 j_0
    =
    \frac{1}
         {1-e^{-D_{\mathcal H}/j_0}}.
\end{equation}

Substituting into $\alpha_{D_T}$ gives
\begin{equation}
    \alpha_{D_T}
    =
    \frac{1-e^{-D_T/j_0}}
         {1-e^{-D_{\mathcal H}/j_0}}.
    \tag{B1}
\end{equation}

For large Hilbert-space dimension
$D_{\mathcal H}\gg j_0$,
$e^{-D_{\mathcal H}/j_0}\ll1$, and therefore
\begin{equation}
    \alpha_{D_T}
    \simeq
    1-e^{-D_T/j_0}.
\end{equation}

Expanding for $\alpha_{D_T}$ close to unity,
\begin{equation}
    \sqrt{\alpha_{D_T}}
    =
    \sqrt{1-e^{-D_T/j_0}}
    \simeq
    1-\frac{1}{2}e^{-D_T/j_0},
\end{equation}
which implies
\begin{equation}
    1-\sqrt{\alpha_{D_T}}
    \simeq
    \frac12 e^{-D_T/j_0}.
\end{equation}
Hence,
\begin{equation}
    \left(1-\sqrt{\alpha_{D_T}}\right)^{1/2}
    \simeq
    \frac{1}{\sqrt2}
    e^{-D_T/(2j_0)}.
    \tag{B2}
\end{equation}

Therefore the BA truncation error decreases exponentially with the
retained basis size $D_T$,
\begin{equation}
    \epsilon_{\rm BA}
    \propto
    e^{-D_T/(2j_0)},
\end{equation}
establishing exponentially fast convergence in the gapped N\'eel
phase.

\subsection*{C2. Critical TLL ($-1<\Delta\le 1$): logarithmic convergence}

\begin{figure}[h]
\includegraphics[width=1.0\columnwidth]{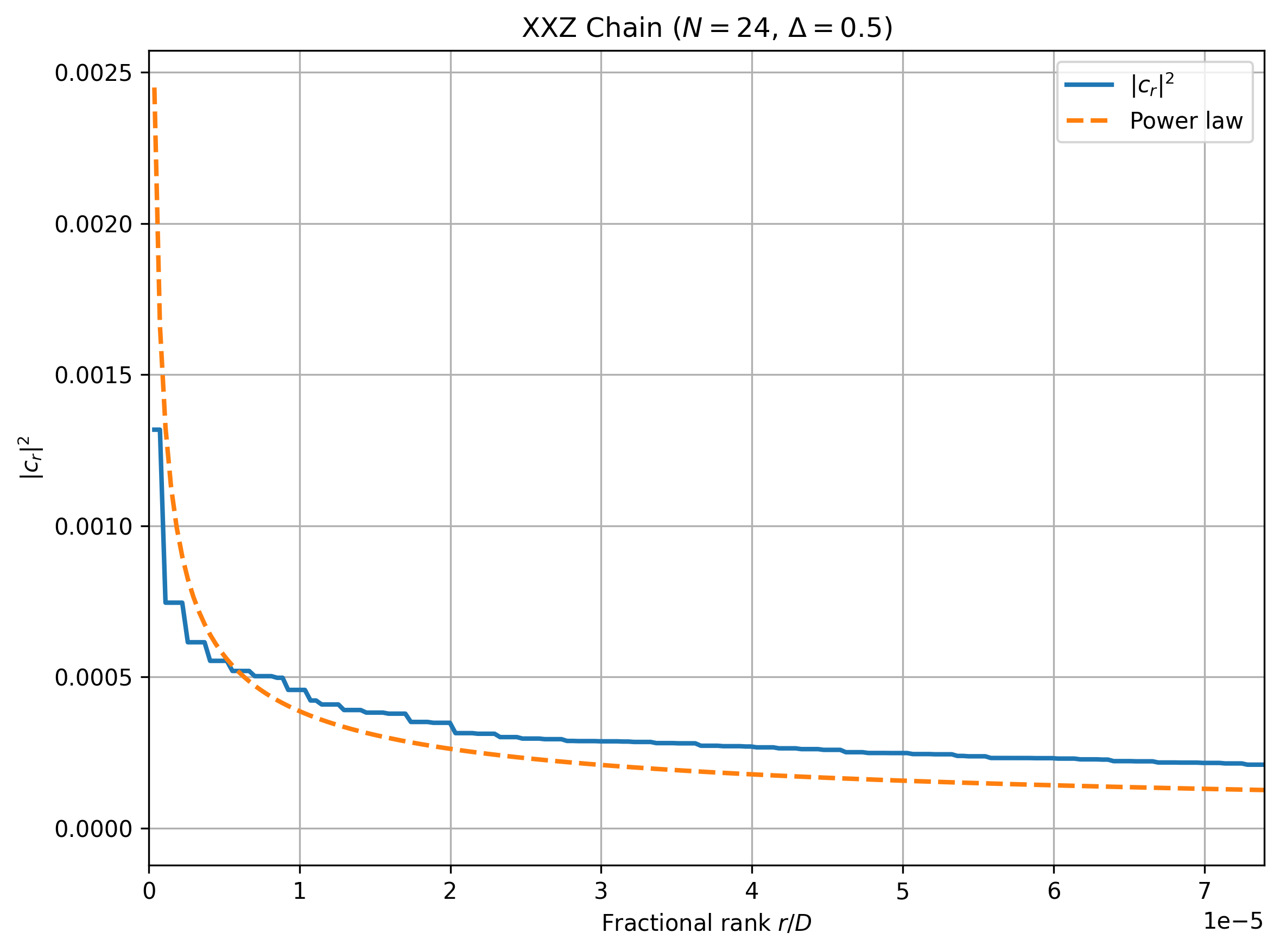}
\caption{For N=24 and $\Delta=0.5$, the computational-basis probabilities $|c_j|^2$ decay is powerlaw with the Hilbert space $j$.}
\label{fig:corsz_comp}
\end{figure}

Let $D_{\mathcal H}$ denote the full Hilbert-space dimension and
$D_T$ the number of basis states retained in the BA subspace.
After sorting the computational-basis probabilities in descending
order,
\begin{equation}
|c_1|^2 \ge |c_2|^2 \ge \cdots,
\end{equation}
we define the retained probability weight
\begin{equation}
\alpha_{D_T}
=
\sum_{j=1}^{D_T}|c_j|^2 .
\end{equation}

In the critical Tomonaga--Luttinger liquid regime
($-1<\Delta\le1$), the sorted probabilities are found
numerically to follow a power-law form,
\begin{equation}
|c_j|^2
\simeq
\frac{\tilde c_0}{j^\gamma},
\label{eq:critical_coeff}
\end{equation}
where $\gamma$ is obtained from fits to the coefficient
distribution.

For large $D_T$, the sum may be approximated by an integral:
\begin{equation}
\alpha_{D_T}
\approx
\int_1^{D_T}
\frac{\tilde c_0}{j^\gamma}\,dj .
\end{equation}

For $\gamma\neq1$,
\begin{equation}
\alpha_{D_T}
=
\frac{\tilde c_0}{1-\gamma}
\left(
D_T^{\,1-\gamma}-1
\right).
\end{equation}

The normalization condition
\begin{equation}
\sum_{j=1}^{D_{\mathcal H}}|c_j|^2 =1
\end{equation}
gives
\begin{equation}
1
=
\int_1^{D_{\mathcal H}}
\frac{\tilde c_0}{j^\gamma}\,dj
=
\frac{\tilde c_0}{1-\gamma}
\left(
D_{\mathcal H}^{\,1-\gamma}-1
\right),
\end{equation}
hence
\begin{equation}
\frac{\tilde c_0}{1-\gamma}
=
\frac{1}
     {D_{\mathcal H}^{\,1-\gamma}-1}.
\end{equation}

Substituting into $\alpha_{D_T}$ yields
\begin{equation}
\alpha_{D_T}
=
\frac{
D_T^{\,1-\gamma}-1
}{
D_{\mathcal H}^{\,1-\gamma}-1
}
=
\frac{
1-D_T^{\,1-\gamma}
}{
1-D_{\mathcal H}^{\,1-\gamma}
}.
\tag{B3}
\end{equation}

\paragraph{Case I: $\gamma>1$.}

For $\gamma>1$, the retained weight is

\begin{equation}
\alpha_{D_T}
=
\frac{
1-D_T^{\,1-\gamma}
}{
1-D_{\mathcal H}^{\,1-\gamma}
}.
\tag{B4}
\end{equation}

Since $D_T\gg1$ and $D_{\mathcal H}\gg1$, both
$D_T^{\,1-\gamma}$ and $D_{\mathcal H}^{\,1-\gamma}$ are small.
Using the binomial expansions

\begin{equation}
(1-x)^{1/2}
=
1-\frac{x}{2}
+\mathcal{O}(x^2),
\end{equation}

and

\begin{equation}
(1-y)^{-1/2}
=
1+\frac{y}{2}
+\mathcal{O}(y^2),
\end{equation}

with
$x=D_T^{\,1-\gamma}$ and
$y=D_{\mathcal H}^{\,1-\gamma}$, we obtain

\begin{align}
\sqrt{\alpha_{D_T}}
&=
\left(
\frac{1-D_T^{\,1-\gamma}}
     {1-D_{\mathcal H}^{\,1-\gamma}}
\right)^{1/2}
\nonumber\\
&\simeq
\left(
1-\frac12 D_T^{\,1-\gamma}
\right)
\left(
1+\frac12 D_{\mathcal H}^{\,1-\gamma}
\right)
\nonumber\\
&\simeq
1
-\frac12 D_T^{\,1-\gamma}
+\frac12 D_{\mathcal H}^{\,1-\gamma}.
\end{align}

Therefore,

\begin{equation}
1-\sqrt{\alpha_{D_T}}
\simeq
\frac12
\left(
D_T^{\,1-\gamma}
-
D_{\mathcal H}^{\,1-\gamma}
\right),
\end{equation}

and hence

\begin{equation}
\left(
1-\sqrt{\alpha_{D_T}}
\right)^{1/2}
\simeq
\frac{1}{\sqrt{2}}
\left(
D_T^{\,1-\gamma}
-
D_{\mathcal H}^{\,1-\gamma}
\right)^{1/2}.
\tag{B5}
\end{equation}

In the physically relevant limit
$D_{\mathcal H}\gg D_T$,

\begin{equation}
D_{\mathcal H}^{\,1-\gamma}
\ll
D_T^{\,1-\gamma},
\end{equation}

so Eq.~(B5) simplifies to

\begin{equation}
\left(
1-\sqrt{\alpha_{D_T}}
\right)^{1/2}
\simeq
\frac{1}{\sqrt{2}}
D_T^{-(\gamma-1)/2}.
\tag{B6}
\end{equation}

Substituting into Eq.~\eqref{eq:master_bound} yields

\begin{equation}
\Delta E
\lesssim
\sqrt{2}\,\|H\|\,
D_T^{-(\gamma-1)/2},
\tag{B7}
\end{equation}

showing that the BA energy error decreases algebraically with the
retained subspace dimension $D_T$ in the critical phase.

\paragraph{Case II: $\gamma<1$.}

For $\gamma<1$, Eq.~(B4) may be rewritten as

\begin{equation}
\alpha_{D_T}
=
\frac{1-D_T^{\,1-\gamma}}
     {1-D_{\mathcal H}^{\,1-\gamma}}
=
\left(
\frac{D_T}{D_{\mathcal H}}
\right)^{1-\gamma}
\frac{
1-D_T^{\,\gamma-1}
}{
1-D_{\mathcal H}^{\,\gamma-1}
}.
\end{equation}

Since $D_{\mathcal H}\gg D_T\gg1$ and $\gamma<1$,
the exponents $\gamma-1$ are negative, implying

\begin{equation}
D_T^{\,\gamma-1}\ll1,
\qquad
D_{\mathcal H}^{\,\gamma-1}\ll1.
\end{equation}

Hence,

\begin{equation}
\alpha_{D_T}
\simeq
\left(
\frac{D_T}{D_{\mathcal H}}
\right)^{1-\gamma}.
\tag{B8}
\end{equation}

Taking the square root gives

\begin{equation}
\sqrt{\alpha_{D_T}}
\simeq
\left(
\frac{D_T}{D_{\mathcal H}}
\right)^{(1-\gamma)/2}.
\end{equation}

Therefore,

\begin{equation}
\left(
1-\sqrt{\alpha_{D_T}}
\right)^{1/2}
=
\left[
1-
\left(
\frac{D_T}{D_{\mathcal H}}
\right)^{(1-\gamma)/2}
\right]^{1/2}.
\tag{B9}
\end{equation}

Substituting into Eq.~\eqref{eq:master_bound} yields

\begin{equation}
\Delta E
\lesssim
2\|H\|
\left[
1-
\left(
\frac{D_T}{D_{\mathcal H}}
\right)^{(1-\gamma)/2}
\right]^{1/2}.
\tag{B10}
\end{equation}

Thus, for $\gamma<1$, the convergence of the BA energy error is
controlled by the ratio $D_T/D_{\mathcal H}$ rather than by a simple
power law in $D_T$ alone.

\paragraph{Marginal case: $\gamma=1$.}

When $\gamma=1$, the integral becomes logarithmic,
\begin{equation}
\alpha_{D_T}
=
\frac{\ln D_T}
     {\ln D_{\mathcal H}}.
\tag{B6}
\end{equation}

Consequently,
\begin{equation}
\left(
1-\sqrt{\alpha_{D_T}}
\right)^{1/2}
=
\left[
1-
\sqrt{
\frac{\ln D_T}
     {\ln D_{\mathcal H}}
}
\right]^{1/2},
\end{equation}
leading to a logarithmically slow convergence of the BA error.

% \subsection*{B3. Required subspace dimensions}

% Setting $\Delta E=\varepsilon$ in the bounds and solving for $L$:
% \begin{align}
%   &\text{Gapped:}\quad
%     L \sim j_0\ln\!\tfrac{2\|H\|}{\varepsilon}
%     = \mathcal{O}\!\left(\mathrm{poly}(D_{\mathcal{H}})\ln\tfrac{1}{\varepsilon}\right),
%   \tag{B6}\\[4pt]
%   &\text{TLL ($\Delta=1$):}\quad
%     L \sim D_{\mathcal{H}}^{\,\|H\|^2/(6\varepsilon^2)},
%   \tag{B7}\\[4pt]
%   &\text{TLL ($0<\Delta<1$):}\quad
%     L \sim D_{\mathcal{H}}^{\,C_K\|H\|^{2/K}/\varepsilon^{2/K}} .
%   \tag{B8}
% \end{align}
% The scaling~(B7) implies that for $\|H\|\sim D_{\mathcal{H}}$ (extensive Hamiltonian),
% the exponent of $D_{\mathcal{H}}$ grows as $D_{\mathcal{H}}^2/\varepsilon^2$, which can become large
% for high-accuracy requirements.  Symmetry filtering effectively
% replaces $D_{\mathcal{H}}$ by the symmetry-sector dimension $N_s<D_{\mathcal{H}}$, reducing the
% prefactor and improving convergence in practice.

\end{document}